\RequirePackage[2018-12-01]{latexrelease}
\documentclass[showpacs,showkeys,11pt,
preprint,preprintnumbers,nofootinbib,
groupedaddress,superscriptaddress,amsmath,amssymb]{revtex4}
\usepackage{amsfonts}
\usepackage{amsmath}
\usepackage{graphicx,subfigure}
\usepackage[export]{adjustbox}
\usepackage{verbatim} 
\usepackage{epsfig}
\usepackage{url}
\usepackage{multirow}
\usepackage{hhline}
\usepackage{feynmp}
 \usepackage{slashed}
 \usepackage{hyperref}
\DeclareUnicodeCharacter{B0}
\usepackage[utf8]{inputenc}
\usepackage[T1]{fontenc}
\usepackage{textcomp}
 \hypersetup{
 colorlinks=true,
 linkcolor=red,
 citecolor=blue,
 }

\newcommand {\be}{\begin{equation}}
\newcommand {\ee}{\end{equation}}
\newcommand {\ba}{\begin{eqnarray}}
\newcommand {\ea}{\end{eqnarray}}

\usepackage{mathtools}

\begin{document}
\title{Magnetic Monopole Phenomenology at Future Hadron Colliders}

\author{Ijaz Ahmed}
\email{ijaz.ahmed@fuuast.edu.pk}
\affiliation{Federal Urdu University of Arts, Science and Technology, Islamabad Pakistan}
\author{Sidra Swalheen}
\email{swalheensidra@gmail.com}
\affiliation{Federal Urdu University of Arts, Science and Technology, Islamabad Pakistan}
\author{Mansoor Ur Rehman}
\email{mansoor@qau.edu.pk}
\affiliation{Department of Physics, Quaid-i-Azam University, Islamabad Pakistan}

\author{Rimsha Tariq}
\email{rimshatariqkbw@gmail.com}
\affiliation{Department of Physics, Quaid-i-Azam University, Islamabad Pakistan}


\begin{abstract}
In the grand tapestry of Physics, the magnetic monopole is a holy grail. Therefore, numerous efforts are underway in search of this hypothetical particle at CMS, ATLAS and MoEDAL experiments of LHC by employing different production mechanisms. The cornerstone of our comprehension of monopoles lies in Dirac's theory which outlines their characteristics and dynamics. Within this theoretical framework, an effective $U(1)$ gauge field theory, derived from conventional models, delineates the interaction between spin magnetically-charged fields and ordinary photons under electric-magnetic dualization. 

The focus of this paper is the production of magnetic monopoles from Drell-Yan and the Photon-Fusion mechanisms to produce velocity-dependent scalar, fermionic, and vector monopoles of spin angular momentum  $0,\frac{1}{2},1$ respectively at LHC. A computational work is performed to compare the monopole pair-production cross-sections for both methods at different center-of-mass energies ($\sqrt{s}$) with various magnetic dipole moments. The comparison of kinematic distributions of monopoles at Parton and reconstructed level are demonstrated including both DY and PF mechanisms.

Extracted results showcase how modern machine-learning techniques can be used to study the production of magnetic monopoles at the Future proton-proton Particle Colliders at 100 TeV.
We demonstrate the observability of magnetic monopoles against the most relevant Standard Model background using multivariate methods such as Boosted Decision
Trees (BDT), Likelihood, and Multilayer Perceptron (MLP). This study compares the performance of these classifiers with traditional cut-based and counting approaches, proving the superiority of our methods.
 
\end{abstract}

\maketitle

\section{Introduction}

Electricity and magnetism have been fundamental in shaping the formation and behavior of diverse entities, ranging from living organisms to celestial bodies like planets, stars, and galaxies. 
The unification of these phenomena into electromagnetism is articulated through Maxwell's equations. This unified framework underscores the significance of electromagnetic duality, elucidated by symmetry arguments.
This duality led to a captivating phenomenon: the existence of magnetic monopoles (MMs), sources of the singular magnetic fields, which emerged as a tantalizing possibility \cite{1,2,3,maxwell}.

Paul Adrien Maurice Dirac offered initial insights into magnetic monopoles (MMs) through his theoretical framework that bridges the conceptual gap between the existence of monopoles and electromagnetic (EM) forces described by Maxwell's equations. 
Dirac's theory suggested that the existence of MMs in any theoretical framework requires a compact $U(1)$ gauge field theory. 
According to Dirac, a monopole represents a point-like structure-less particle serving as the magnetic analog of electric charge, thus providing the basis for the quantization of electric charge. These ground-breaking revelations are encapsulated in Dirac's quantization condition (expressed in natural SI units):
\begin{equation}
q_{e} \times {g} =2\pi n,
    \qquad\qquad n\in\mathbb{Z} 
\end{equation}
where `$q_e$' and `$g$' represent electric and magnetic charges, respectively, with `$n$' as an integer. Thus Dirac's theory intriguingly creates space for the existence of magnetic monopoles; he famously remarked, ``One would be surprised if nature had made no use of it'' \cite{Dirac31,Dirac48}.

The minimum conceivable magnetic charge, often referred to as the Dirac charge ($g_D$), is expressed as
\begin{equation}
g_{D} = \frac{1}{2\alpha}\,e \approx 68.5 \, e,
\end{equation}
where `$\alpha = e^2/4\pi \approx 1/137$' denotes the fine-structure constant. Although monopoles formally symmetrize Maxwell's equations, there arises an asymmetry due to the Dirac quantization condition as the minimum value of magnetic charge significantly exceeds the smallest electric charge.
Comparatively, the dimensionless magnetic coupling constant, $\alpha_g = g^2_D/4\pi \approx 34.25$, is significantly greater than one. As a result, perturbative calculations involving monopoles encounter substantial challenges. Moreover, it is not possible to estimate the mass and spin of the magnetic monopole within Dirac's model. Particularly, the lack of knowledge regarding the monopole's mass is a crucial missing element for evaluating the feasibility of producing such objects at colliders.

ls
Besides these point-like MMs, which lack both gauge and Lorentz invariance, several non-Abelian gauge theories also suggest the existence of composite soliton-like monopoles. Specifically, they are ineluctable in Grand Unified Theories (GUTs). They are expected to have masses around the spontaneous symmetry breaking scale, which for GUTs typically fall in the $10^{16}$ GeV or higher range. 
These GUT mass monopoles, originating from the early epoch, certainly lie outside the reach of the current particle colliders while generating flux levels surpassing the Parker limit. Nevertheless, the density of these monopoles could be diluted through the mechanisms of inflation. For recent work, see \cite{Rehman:2008qs,Khalil:2010cp,Civiletti:2011qg,Rehman:2014rpa,Rehman:2018gnr,Masoud:2019gxx,Lazarides:2020zof,Ahmed:2023rky,Afzal:2023cyp,Ijaz:2023cvc,Ijaz:2024zma,Zubair:2024quc,Ahmed:2024iyd} for dilution of monopole density in GUTs such as $SO(10)$, $SU(5)$, $\chi SU(5)$, $SU(4)_c \times SU(2)_L \times SU(2)_R$ and $SU(4)_c \times SU(2)_L \times U(1)_R$.

The potential existence of intermediate-mass monopoles, which could survive through the inflationary epoch, is explored in \cite{Lazarides:1980cc,Senoguz:2015lba,Lazarides:2019xai,Chakrabortty:2020otp,Maji:2022jzu,Lazarides:2023iim}. This could lead to an observable flux of primordial magnetic monopoles, potentially approaching or falling within a few orders of magnitude below the Parker limit. Recent insights from \cite{Lazarides:2021tua,Raut:2022ryj} indicate that unlike the superheavy GUT monopoles, the trinification monopole, due to its lighter nature, could potentially be detectable at high-energy colliders.
Interestingly, MMs could also emerge from electroweak symmetry breaking with a few TeV masses \cite{tHooft,Polyakov,nambu,maison,Hung,Ellis:2020bpy,Benes:2024smx}. For a recent realization of a composite monopole in the standard model that only carries EM magnetic flux, see \cite{Lazarides}. These theoretical frameworks suggest the possibility of generating magnetic monopoles (MMs) at LHC colliders, renewing interest in their investigation, despite the absence of any observed isolated magnetic monopoles thus far \cite{MoEDAL21, MoEDAL24,MoEDAL:2019ort,ATLAS}.

\begin{figure}[h]
 \centering
    \subfigure[]{%
    \includegraphics[width=0.4\textwidth]{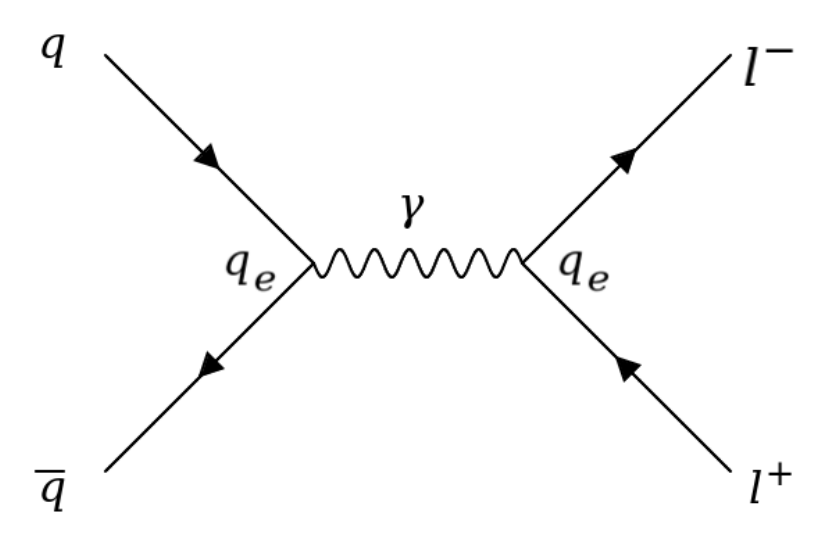}\label{l}}
    \qquad
\centering
    \subfigure[]{%
    \includegraphics[width=0.4\textwidth]{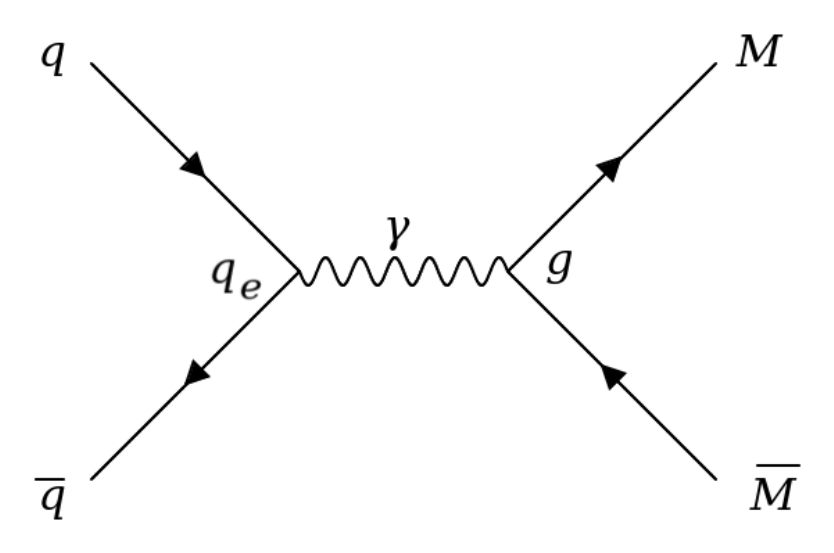}\label{m}}
    \caption{Feynmann diagrams of Drell-Yan (DY) mechanism.}
    \label{DY}
\end{figure}

The LHC colliders utilize Drell-Yan (DY), Photon-Fusion (PF), and Schwinger mechanisms for monopole production and have set limits on the charge and mass of monopoles (for details see \cite{MoEDAL1604,MoEDAL1611,MoEDAL17,MoEDAL:2019ort,MoEDAL21,MoEDAL23,MoEDAL24,ATLAS12,ATLAS15,ATLAS}). However, our subject of interest here is DY and PF mechanisms as they are commonly used. For the DY process, the production of a virtual photon or $Z$ boson takes place via the annihilation of quarks sourced from different hadrons \cite{Drell}. Within the Standard Model (SM), this photon decays into a lepton-antilepton pair facilitating lepton synthesis from the exchange of electric charges `$q_e$' as depicted in Fig.~\autoref{l}. While for MM production shown in Fig.~\autoref{m}, the virtual photon subsequently decays into a monopole-antimonopole pair, leading to MM synthesis through magnetic charge `$g$'.

\begin{figure}[h]   
 \centering 
    \subfigure[\,Three vertex diagram for spin
    $0,\frac{1}{2},1$]{\includegraphics[width=0.4\textwidth]{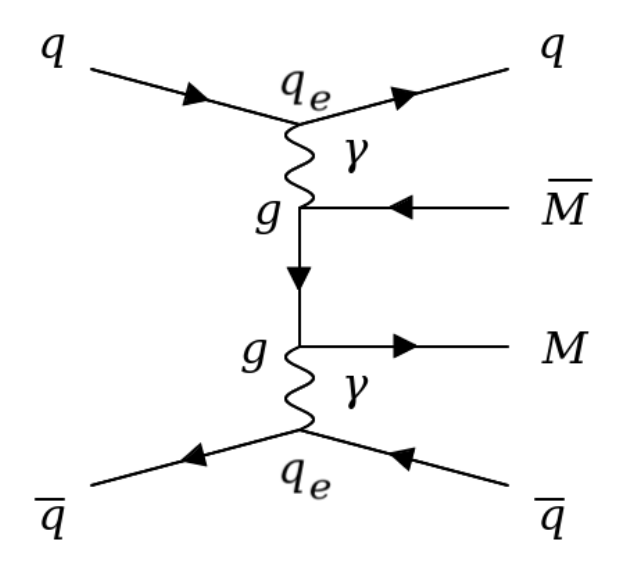}}
\centering
    \subfigure[\,Four vertex diagram for spin $0,1$]{ \includegraphics[width=0.4\textwidth]{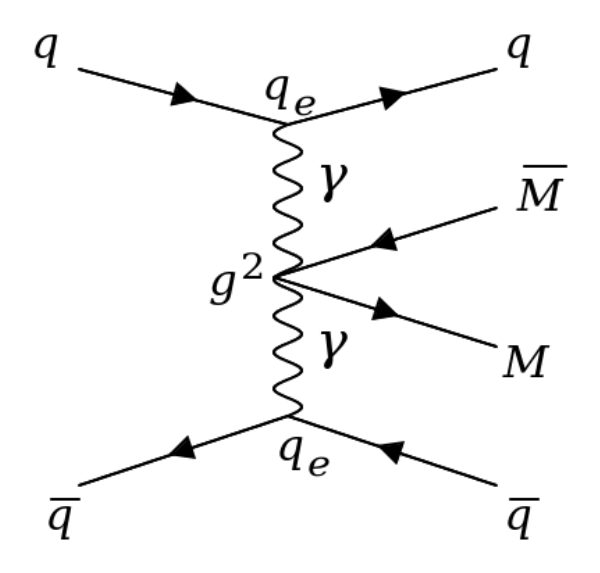}}
    \caption{Feynmann diagrams of Photon-Fusion (PF) mechanism for monopole synthesis.} 
    \label{PF}
\end{figure}

 Let's now switch our focus to the PF mechanism, which found its roots in 1937 amidst investigations into high-energy particle collisions. This mechanism involves the collision of high-energy photons, typically sourced from gamma-ray emissions, triggering the formation of particle-antiparticle pairs such as electron-positron pairs \cite{Breit}. This process occurs through the interaction of the photons with the intense electromagnetic fields surrounding massive celestial objects like black holes or neutron stars, rendering it crucial for analyzing various astrophysical phenomena. Additionally, it's fascinating that this mechanism also harbors the potential to generate monopole-antimonopole pairs, as illustrated in \autoref{PF}.
 
 When investigating the creation of monopole-antimonopole pairs through PF and DY mechanisms, or exploring the propagation of monopoles in matter used for their detection and capture, analyzing kinematic distributions analytically proves to be highly beneficial. One of the key phenomenological parameters that demands our attention is the relative velocity of the monopole. This parameter, often termed the ‘coupling boost’, gains particular importance due to the high coupling strength of MMs and is defined as
\begin{equation}
    \beta = \sqrt{1 - \frac{4M^{2}}{\hat{s}}}, \label{E1}
\end{equation}
where Mandelstam variable `$\hat{s}$' further connotes dependency on the Center-of-Mass Energy (CME) of the incoming particles `$\sqrt{s}$' expressed as
\begin{equation}
 \sqrt{s} = 2E_{\gamma/q}.
\end{equation}

In \cite{Baines:2018ltl}, previous Drell-Yan and Photon-Fusion studies \cite{Kurochkin:2006jr,Dougall:2007tt} are performed for $S-1/2$ that was extended for $S-0$ and $S-1$. Recent MoEDAL experiments \cite{MoEDAL:2019ort} use the Photon-Fusion results of \cite{Baines:2018ltl} that are also presented here. Here, it's worth noting that \textsc{MadGraph} employs Heaviside–Lorentz units instead of SI units. In this system, the electromagnetic vertex is represented as:
\begin{equation}
C_{EM} = q_{e},
\end{equation}
where $q_e=\sqrt{4\pi\alpha}$. Conversely, the vertex for the coupling between photons and monopoles is given by
\begin{equation}
C_{MM} = g.
\end{equation}
The Dirac charge can now be formulated as
\begin{equation}
g_D = \sqrt{\frac{\pi}{\alpha}}.
\end{equation}
For the velocity-dependent coupling of monopoles, the value of $g_D$ becomes $\beta \sqrt{\pi/\alpha}$, where $\beta$ is defined in \autoref{E1}.

This paper will also cover discussions on arbitrary values of another important phenomenological parameter named `magnetic dipole moment' represented by $k$. In the limit of large $k$, despite the significant magnetic coupling, the cross sections for monopole-pair production may remain finite. This scenario makes it reliable to use the Feynman-like diagrams within the effective theory. Perturbativity conditions, as mentioned in \cite{Baines:2018ltl}, particularly concerning slowly moving monopoles, hold relevance for MoEDAL searches \cite{MoEDAL014}.

The paper is structured as follows: Section II delves into the cross-section analysis of monopole pair production via PF and DY processes, considering spins $0$, $\frac{1}{2}$, and $1$. Section III explores monopole phenomenology, kinematic distributions, and the \textsc{MadGraph} implementation. Section IV presents a multivariate analysis of magnetic monopoles by selecting three benchmark points for numerical values. The conclusion, comprising key findings and implications, is presented in Section V.

\section{Cross-sections for monopole pair production}\label{csm}

We discuss the electromagnetic interaction of a monopole of spin $S\,=\,0,\,\frac{1}{2},\,1$ with a photon.
The corresponding theory is an effective $U(1)$ gauge theory obtained after appropriate dualization of the pertinent field theories describing the interactions of charged spin-S fields with photons. The analytical expressions of the differential cross section for spin $S\,=\,0,\,\frac{1}{2},\,1$ monopole-antimonopole production can be found in \cite{Baines:2018ltl}.

\subsection{Scalar Monopoles}

A spin-$0$ monopole interacting with a massless gauge field representing the photon will be discussed here. The Lagrangian that describes the electromagnetic interactions of the monopole is given by dualization of the Scalar Quantum Electrodynamics (SQED) Lagrangian \cite{Mitsou:2019aut},
\begin{equation}
    \mathcal{L}= - \frac{1}{4}F^{\mu v}F_{\mu v}+(D^{\mu}\phi)^{\dag}(D_{\mu}\phi)-M^2\phi^{\dag}\phi ,
\end{equation}
where `$\dag$' is Hermition cojugate, $D_\mu = \partial_\mu - ig(\beta)\mathcal{A}_\mu, \mathcal{A}_\mu$ is photon field with field tensor $F_{\mu v}=\partial_\mu \mathcal{A}_v - \partial_v\mathcal{A}_\mu$ and `$\phi$' is scalar monopole field.

\begin{figure}[h]  
    \centering
     \subfigure[\,DY process]{%
     \includegraphics[width=7cm,height=5.5cm]{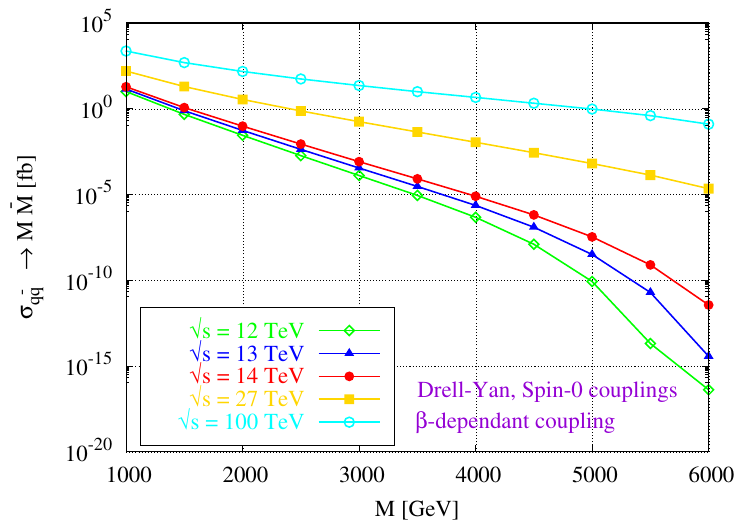}}
\centering
  \subfigure[\,PF process]{%
     \includegraphics[width=7cm,height=5.5cm]{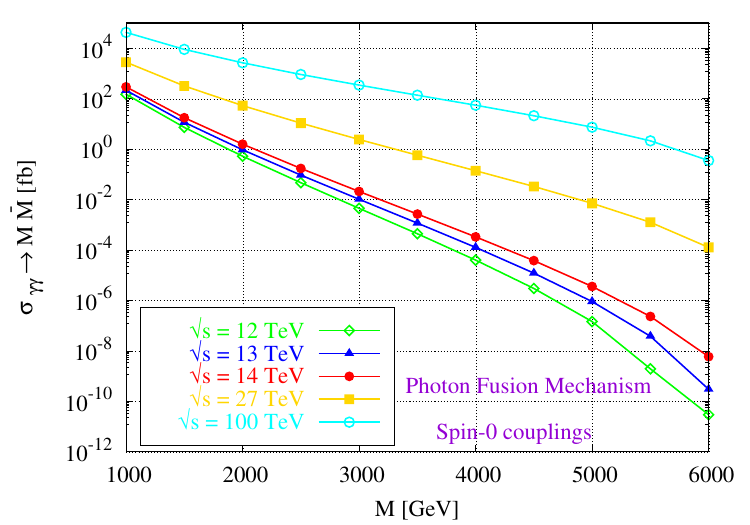}}

 \caption{This plot focuses on the relationship between cross-section fluctuations, mass, and various CMEs for spin-1 monopoles selected for DY mechanism (left) and PF mechanism (right) providing description for exisiting and future colliders. }
 \label{eee}
 \end{figure}

The cross-sections for the PF and DY mechanisms at CMEs of $\sqrt{s}$= 12 TeV, 13 TeV, 14 TeV, 27 TeV and 100 TeV are plotted against the monopole mass for spin-$0$ monopoles. The analysis, depicted in \autoref{eee}, utilized the LUXqed and NNPDF23 PDFs for PF and DY processes respectively while disregarding the magnetic moment term for scalar monopoles with masses ranging from $1$ TeV to $6$ TeV. The production cross-section of a scalar monopole is the probability of producing a scalar monopole in a given interaction or process. It depends on the energy of the colliding particles, the interaction type, and the mass of the scalar monopole. As the mass of a scalar monopole increases, the production cross-section generally decreases. This is because, in high-energy particle collisions, the CME determines the maximum possible mass that can be produced. If the mass of the scalar monopole is too high compared to the available CME, it becomes less likely to be produced and vice versa.

 \subsection{Fermionic Monopoles}

 The unknown origin of monopole magnetic moment led us to conclude that it is only generated through different quantum spin interactions. The effective Lagrangian for the spinor-monopole-photon interactions takes the following form,
\begin{equation}
    \mathcal{L} = - \frac{1}{4}F^{\mu v}F_{\mu v}+\overline{\psi}(i\slashed{D}-m)\psi - k g(\beta)  F_{\mu v} \overline{\psi}\,\sigma^{\mu \nu}\psi ,
    \label{4.6}
\end{equation}
where `$F_{\mu v}$' is electromagnetic field strength tensor, `$k$' denotes magnetic moment, $\sigma^{\mu \nu} =  [\gamma^{\mu} , \gamma^v]/4$  is a commutator of Dirac's $\gamma$ -matrices, and $\slashed{D}=\gamma^\mu [\partial_{\mu}-ig(\beta)\mathcal{A}_\mu ]$ is the covariant derivative. The magnetic coupling `$g(\beta)$' is at most linearly dependent on the velocity of monopole `$\beta$'.

\begin{figure}[h]   
\centering
     \subfigure[\,DY process]{%
   \includegraphics[width=7cm,height=5.5cm]{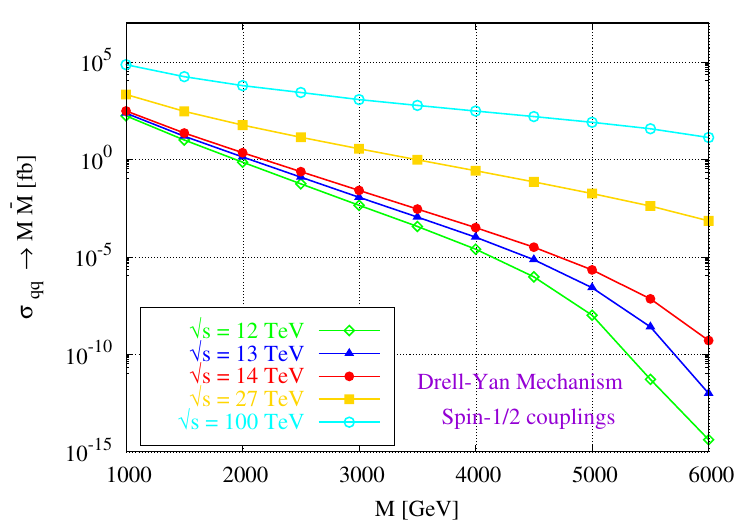}\label{L1}}
     \qquad
 \centering
    \subfigure[\,PF process]{%
   \includegraphics[width=7cm,height=5.5cm]{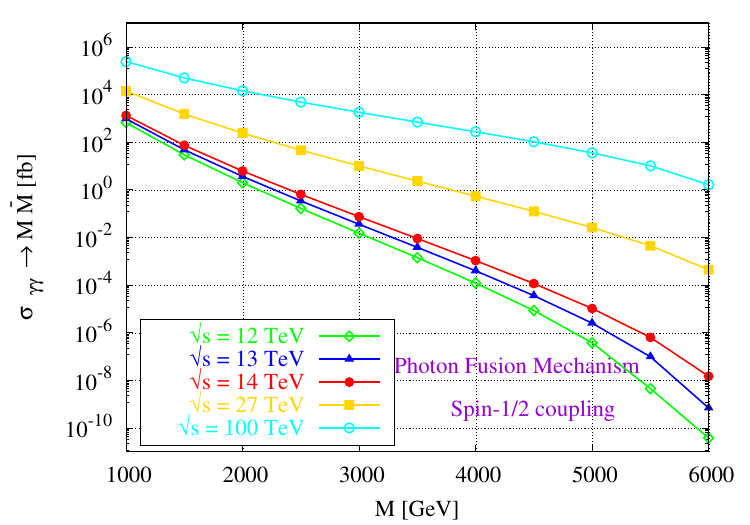}\label{L2}}

 \caption{The cross-section fluctuations for monopoles with spin $S-\frac{1}{2}$ (spin-half monopoles) as a function of their mass across different values of CMEs by choosing DY mechanism (left) and PF mechanism (right)}
 \label{abcd}
 \end{figure}

\begin{figure}
    \centering
     \subfigure[\,DY process]{%
   \includegraphics[width=7cm,height=5.5cm]{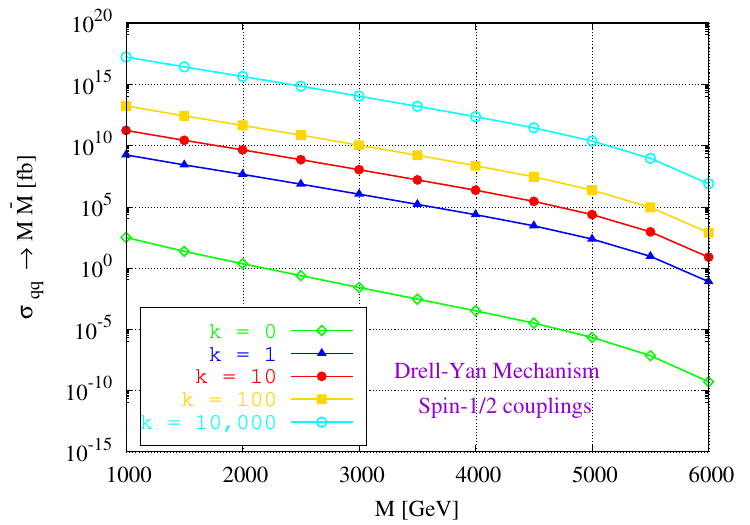}\label{R1}}
     \qquad
     \centering
     \subfigure[\,PF process]{%
   \includegraphics[width=7cm,height=5.5cm]{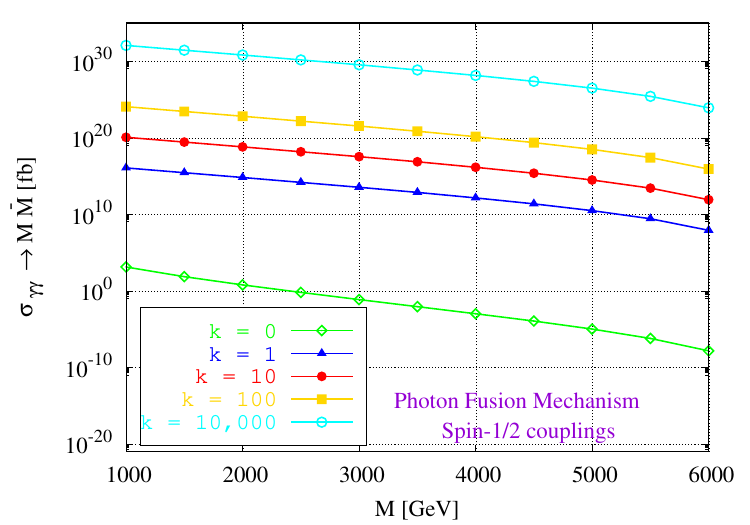}\label{R2}}
    \caption{The cross-section fluctuations for monopoles with spin $S-\frac{1}{2}$ (spin-half monopoles) as a function of their mass across different values of the parameter `$k$'  by adopting DY mechanism (left) and PF mechanism (right)}
    \label{abcd1}
\end{figure}

Here we discuss the total cross-section for the processes $\gamma \gamma \to M \bar{M}$ and $q \bar{q} \to M \bar{M}$, where $M$ and $\bar{M}$ represent monopole and anti-monopole respectively. The monopole mass range considered is between 1 TeV and 6 TeV, and the analysis is performed for different values of the parameter $k$ (0, 1, 10, 100, 10000). The choice of these specific $k$ values is based on the relatively low uncertainty in the photon distribution function in the proton, which is derived from the LUXqed model for the photon-hadron (PF) interaction. This decision helps improve the accuracy of the calculations. The cross-section is determined at various $k$ values for both PF and Drell-Yan (DY) processes at a center-of-mass energy of 14 TeV. The center-of-mass energy is a crucial parameter in particle collisions, as it determines the available energy for interactions. The results indicate that the value of $\sigma$' increases with the value of the $k$ parameter. Despite this, it is essential to note that the $\sigma$' for non-zero $k$ values remains finite, implying that the interactions under study are still observable and measurable in experiments. The data related to these findings can be found in the referenced figures, \autoref{abcd} and \autoref{abcd1}.

\subsection{Vector Monopoles}

For the first time, S-$1$ monopoles have been examined recently by MoEDAL experiment for the Drell-Yan process \cite{MoEDAL17}. The monopole with spin-$1$ is referred to as massive vector meson $W_\mu$ which interacts only with massless gauge field $\mathcal{A}_\mu$ within the framework of gauge invariant Proca field theory. Not having any underlying theory for point-like monopole a $k$ term is included in Lagrangian as the free phenomenological parameter.

\begin{figure}[h]   
    \centering
     \begin{subfigure}
     {\includegraphics[width=7cm,height=5cm]{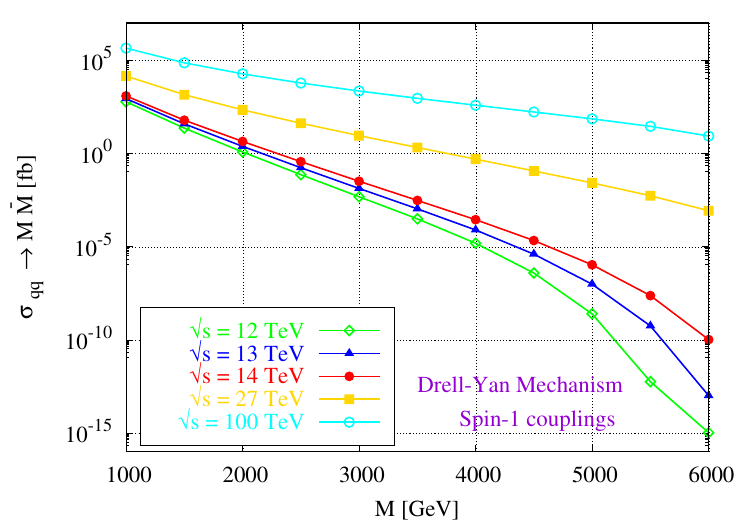}}
 \end{subfigure}
  \begin{subfigure}
     {\includegraphics[width=7cm,height=5cm]{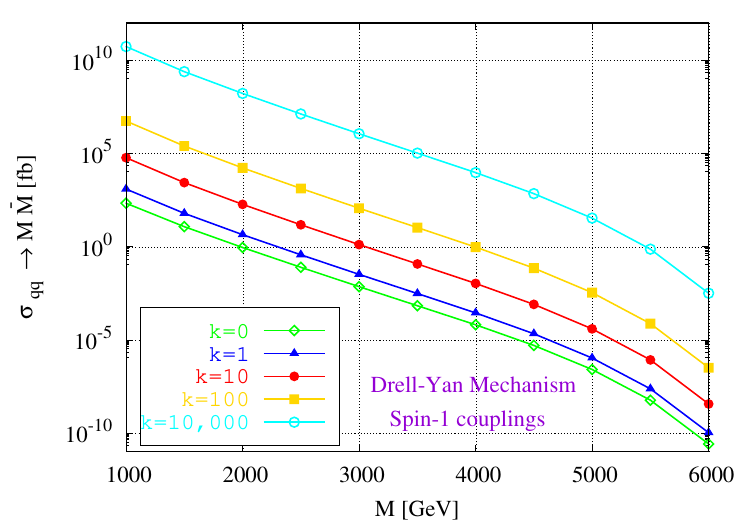}}
 \end{subfigure}
 \begin{subfigure}
     {\includegraphics[width=7cm,height=5cm]{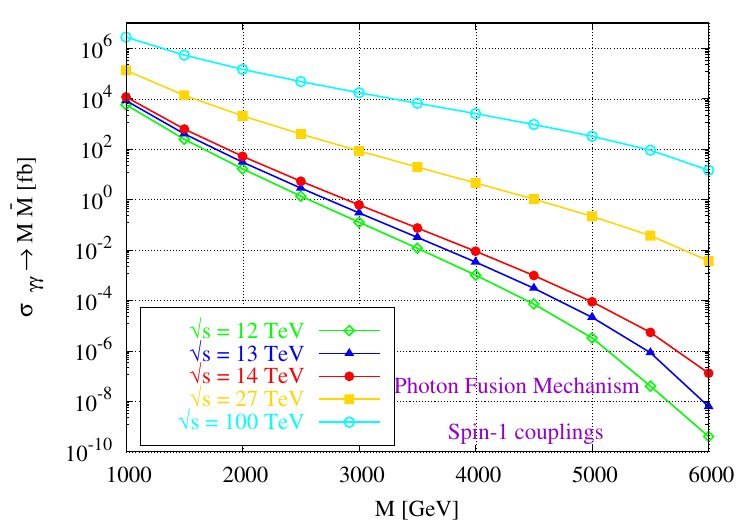}}
 \end{subfigure}
 \begin{subfigure}
     {\includegraphics[width=7cm,height=5cm]{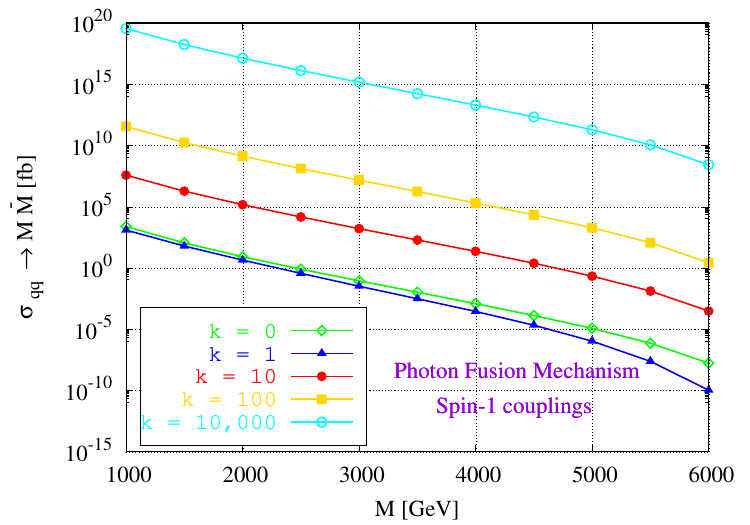}}
 \end{subfigure} 
 \caption{Total cross-section for the different center of mass energy for spin-1 monopoles through Photon fusion process for magnetic moment term $k$ = 1, Cross section for different values of k for spin-1 monopoles at mass range $M$ from 1 TeV - 6 TeV and $\sqrt{s_{\gamma\gamma}}$ = 14 TeV}
 \label{ddd}
 \end{figure} 

The effective Lagrangian acquired by imposition of electric-magnetic duality on Lagrangian's effective term for interaction of $W^{\pm}$ gauge bosons with the photon in SM context takes the form:
\begin{equation}
    \mathcal{L}=- \xi(\partial_{\mu}W^{\dag\mu})(\partial_{\nu} W^\nu)-\frac{1}{2}(\partial_{\mu}\mathcal{A}_v)  (\partial^{v}\mathcal{A}^{\mu}) - \frac{1}{2}G^{\dag}_{\mu v}G^{\mu v}-M^2 W_{\mu}^{\dag}W^{\mu}-ig(\beta) k F^{\mu v}W_{\mu}^{\dag}W_v,
\end{equation}
where the parameter $\xi$ is a
gauge-fixing parameter and $G^{\mu v}=D^{\mu}W^{v}-D^vW^\mu$, $D_\mu = \partial_\mu -ig(\beta)\mathcal{A}_\mu$.\\
 Monopole mass $M$ = 1 TeV - 6 TeV is chosen for various values of $k$ at $\sqrt{s}=14$ TeV graphs for these cross sections are given in \autoref{ddd}.

 \subsection{Comparison}

\begin{figure}[h]  
    \centering
  \begin{subfigure}
     {\includegraphics[width=7cm,height=6cm]{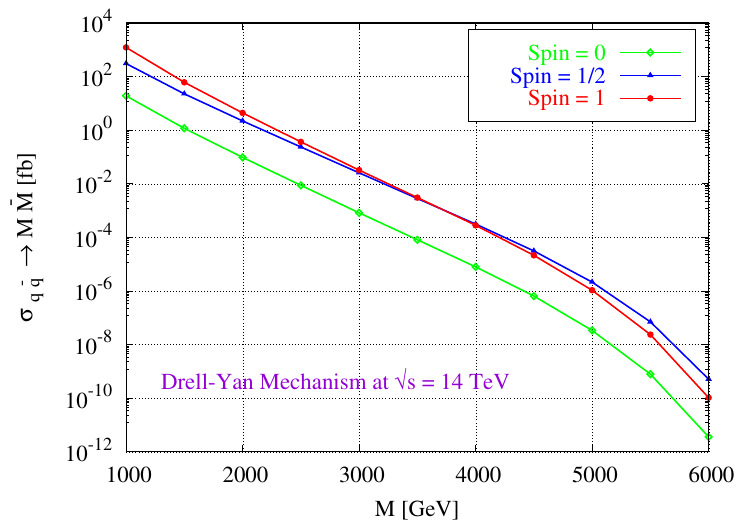}}
 \end{subfigure}
  \begin{subfigure}
     {\includegraphics[width=7cm,height=6cm]{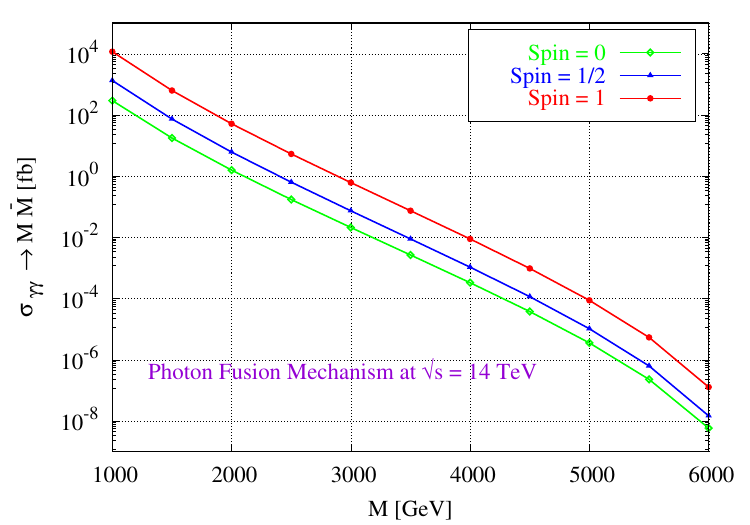}}
 \end{subfigure}
  \caption{Comparison of total cross-section via Drell-Yan and Photon Fusion with ${k}=0$ for spin-1/2 and ${k}=1$ for spin-1  at $\sqrt{s}=14$ TeV, verses mass are drawn, while no $k$ term is included in spin-0 monopoles.}
  \label{mix1}
\end{figure}
 
 As seen in the \autoref{mix1} graph demonstrates the behavior of the monopole's cross-section formation as a function of its mass for different spins (0, 1/2, and 1) in the DY mechanism (left) and PF mechanism (right). The LHC energy is set at 14 TeV. Throughout the entire mass range of 1 TeV to 6 TeV, spin-1/2 and spin-1 monopoles exhibit higher cross-sections than spin-0 monopoles. However, it is crucial to note that for spin-1/2 and spin-1 monopoles, the spin-1 variant dominated up to a mass of 3500 GeV, while spin-1/2 took over at masses greater than or equal to 3500 GeV. This observation implies that, depending on the monopole's mass, different spins may have varying levels of interaction cross-sections at the LHC energy of 14 TeV. Reduction of cross section is seen at higher masses of monopoles as a general trend. Understanding these relationships can help guide experimental efforts and improve our understanding of monopole physics.
 
\begin{figure}[h]  
    \centering
  \subfigure[\,Spin-$0$ monopoles]{\includegraphics[width=5.3cm, height=5cm]{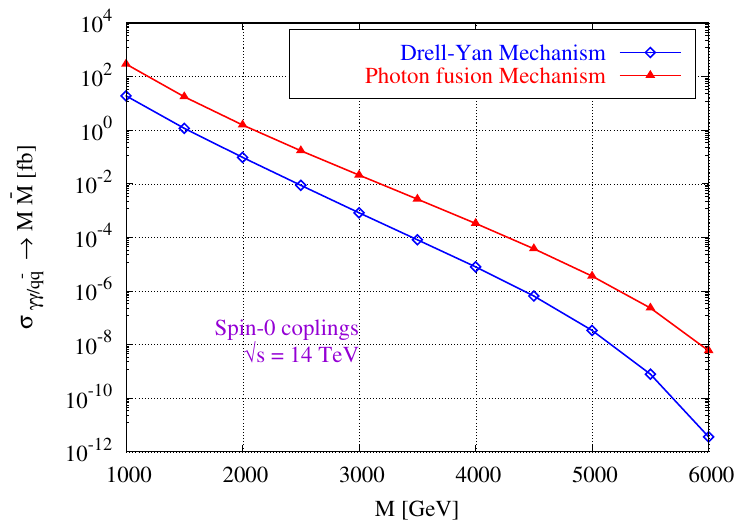}}
 \subfigure[\,Spin $\frac{1}{2}$ monopoles]{\includegraphics[width=5.3cm, height=5cm]{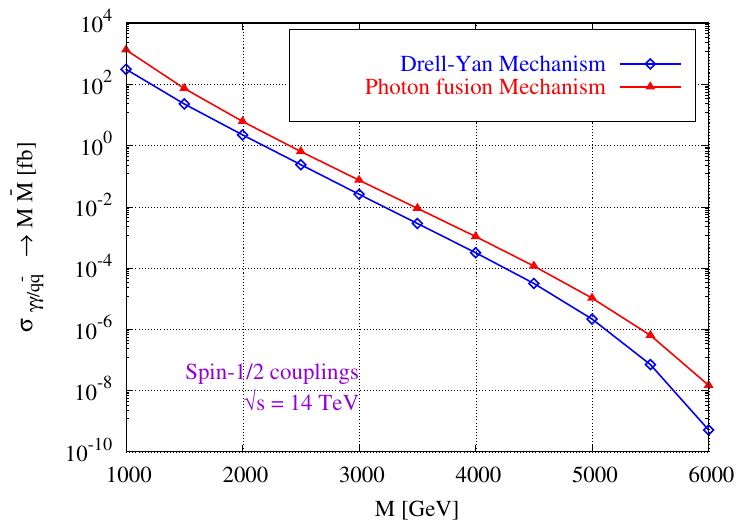}}
  \subfigure[\,Spin-$1$ monopoles]{\includegraphics[width=5.3cm, height=5cm]{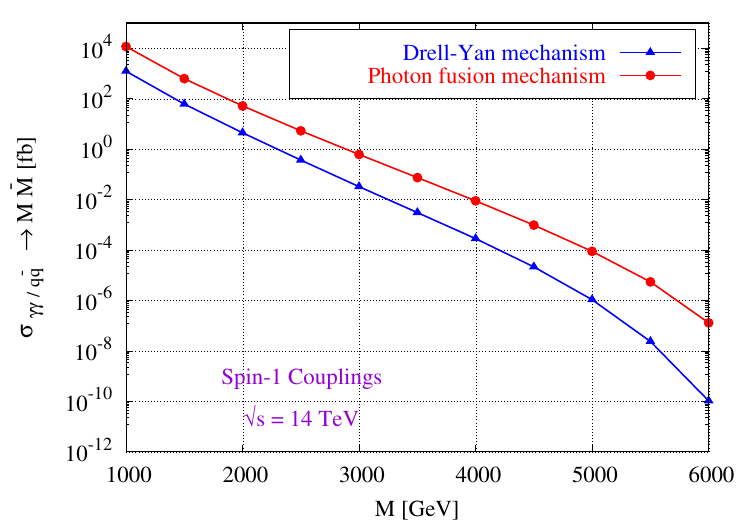}}
    \caption{Comparison of `$\sigma$' for DY and PF processes at $\sqrt{s}=14$ TeV.}
    \label{mix}
\end{figure}

A comparison of both processes at $\sqrt{s}$ = 14 TeV is shown in \autoref{mix}.


\section{Monopole phenomenology and \textsc{MadGraph} implementation}

Magnetic monopoles have long intrigued physicists due to their potential to explain several phenomena in our universe. The study of these entities, known as monopole phenomenology, involves investigating their properties, interactions, and potential effects on the universe. The \textsc{MadGraph} generator \cite{Alwall}, a powerful computational tool, plays a pivotal role in this exploration. It allows for the simulation of particle interactions, including the creation of MMs (for detail see \cite{Baines:2018ltl}). By implementing the UFO (Universal Feynman Output) model within the \textsc{MadGraph} \cite{Degrande}, one can simulate various monopole production mechanisms and analyze the resulting kinematic distributions. It covers photon-monopole coupling that is both independent and dependent on $\beta$ with the incorporation of three distinct spin cases: $0$, $\frac{1}{2}$, and $1$.

\textsc{FeynRules} \cite{Alloul}, a \textsc{Mathematica} interface, is used to build the UFO model from the Lagrangian. In this process, a text file is populated with the model's parameters, such as mass, spin, electric and magnetic charges, coupling constant, and the type of field (fermionic or bosonic), alongside their respective Lagrangian. \textsc{FeynRules} then uses this data to create the UFO model. Within the generated UFO model, the monopole's velocity $\beta$, as delineated in \autoref{E1}, is incorporated as a form factor. Formulating the form factor accurately is ensured by adhering to the guidelines in \cite{FF}.

Previously, the DY process for monopoles had been integrated both for $\beta$ dependent and independent coupling into \textsc{MadGraph} via a \textsc{Fortran} code for spin-$0$ and spin-$\frac{1}{2}$ monopoles, which ATLAS \cite{ATLAS12, ATLAS15} and MoEDAL \cite{MoEDAL1604,MoEDAL1611,MoEDAL17} have employed to interpret their search results. In the scope of the current study, these original \textsc{Fortran} setups were reformulated into UFO models, following their PF counterparts, and then augmented to incorporate the spin-$1$ case. This advancement was crucial in the latest search for MMs conducted by MoEDAL collaboration \cite{MoEDAL17}, providing a more robust framework for analysis.

Simulations were performed to evaluate the cross-sections of both PF and DY mechanisms across all three distinct spin scenarios utilizing the \textsc{MadGraph} UFO model. These simulations were conducted at a center-of-mass energy of $14$ TeV, within the mass range of $1 - 6$ TeV. The results, showcased in \autoref{mix}, reveal an absence of interference between the PF and DY processes at the tree level. The PF mechanism exhibits dominance at the LHC’s energy scale of $14$ TeV across the entire mass range of interest and for all spin configurations. These findings emphasize the significance of the PF production mechanism in LHC experiments, while still acknowledging the relevance of the DY process.


\subsection{Monopole phenomenology}

In collider experiments, not only cross-sections of generated monopoles but also, the study of their angular distributions is crucial, as it enhances the understanding of monopole searches. This significance stems from the fact that detectors' geometrical acceptance and efficiency vary across the solid angle surrounding the interaction point. When examining kinematic distributions for $\gamma\gamma$ and $q\Bar{q}$ scattering, results from UFO models align well with the observed transverse momentum ($p_T$), transverse energy ($E_T$), and relative velocity ($\beta$) of MMs.

In this study, we conduct a comparative analysis of the kinematic distributions for Photon-Fusion ($\gamma\gamma$) and Drell-Yan ($q\bar{q}$) processes. The analysis utilizes a $\beta$-dependent UFO model within the \textsc{MadGraph} framework. This refined approach ensures a more accurate interpretation of experimental data, facilitating the advancement of monopole research. 

We have simulated monopole-antimonopole pair production for both DY and PF processes at $\sqrt{s} = 14$ TeV. For the DY process, the NNPDF23 at \textbf{LO} is employed \cite{Ball}, while the LUXqed \cite{Manohar} is selected for the PF process due to its minimal uncertainty in photon distribution within the proton \cite{Manohar17}. The simulation is configured for $30,000$ events, with the monopoles assigned a magnetic charge of $1g_D$. It is important to note that the magnetic charge value does not influence the kinematic spectra. For ease of comparison between the shapes of the distributions, normalization is applied to equalize the event count across the distributions.

\begin{figure}[h] 
\centering
\subfigure[\,Scalar monopoles]{%
\includegraphics[width=5.2cm,height=5cm]{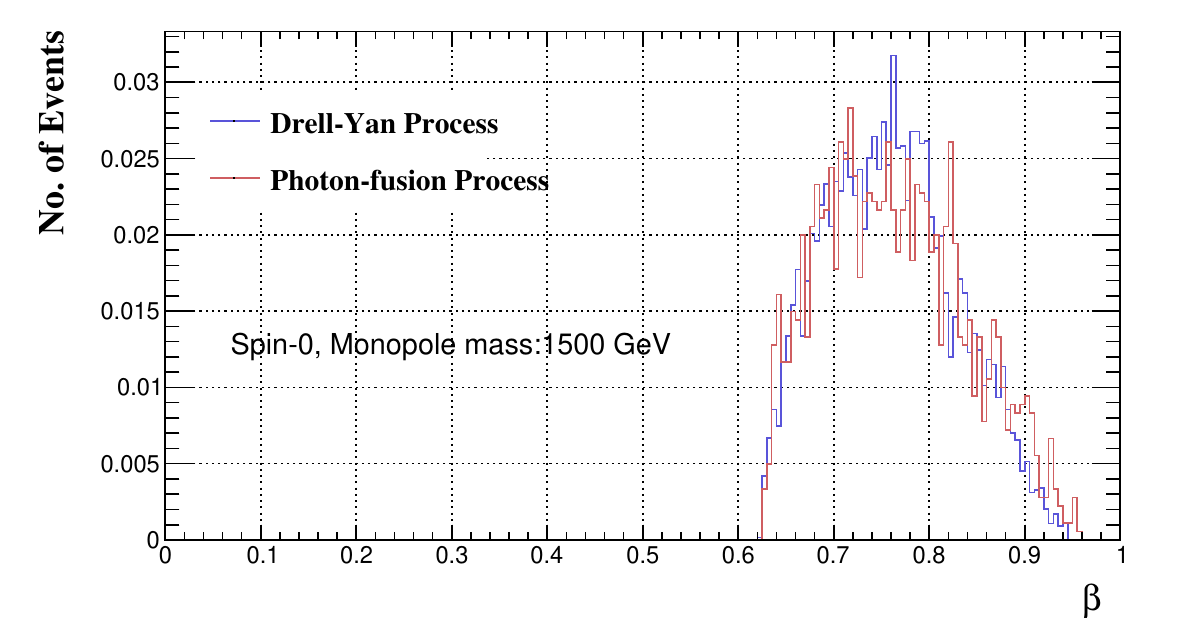}\label{l1}}
\centering
 \subfigure[\,Fermionic monopoles]{%
\includegraphics[width=5.2cm,height=5cm]{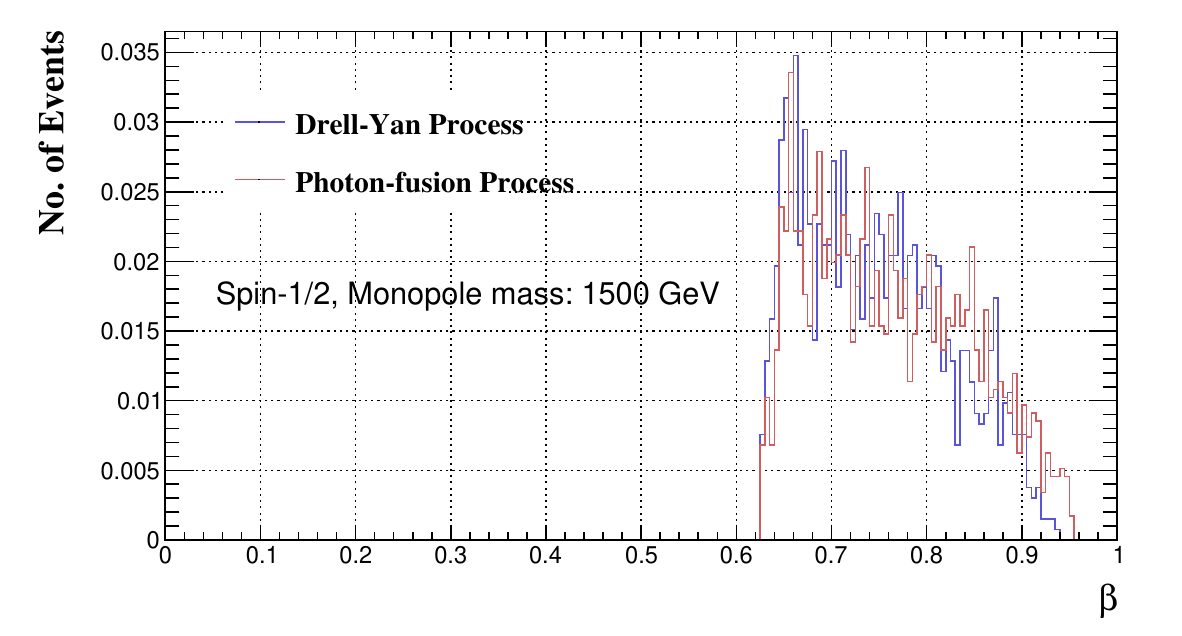}\label{c1}}
\centering
 \subfigure[\,Vector monopoles]{%
 \includegraphics[width=5.2cm,height=5cm]{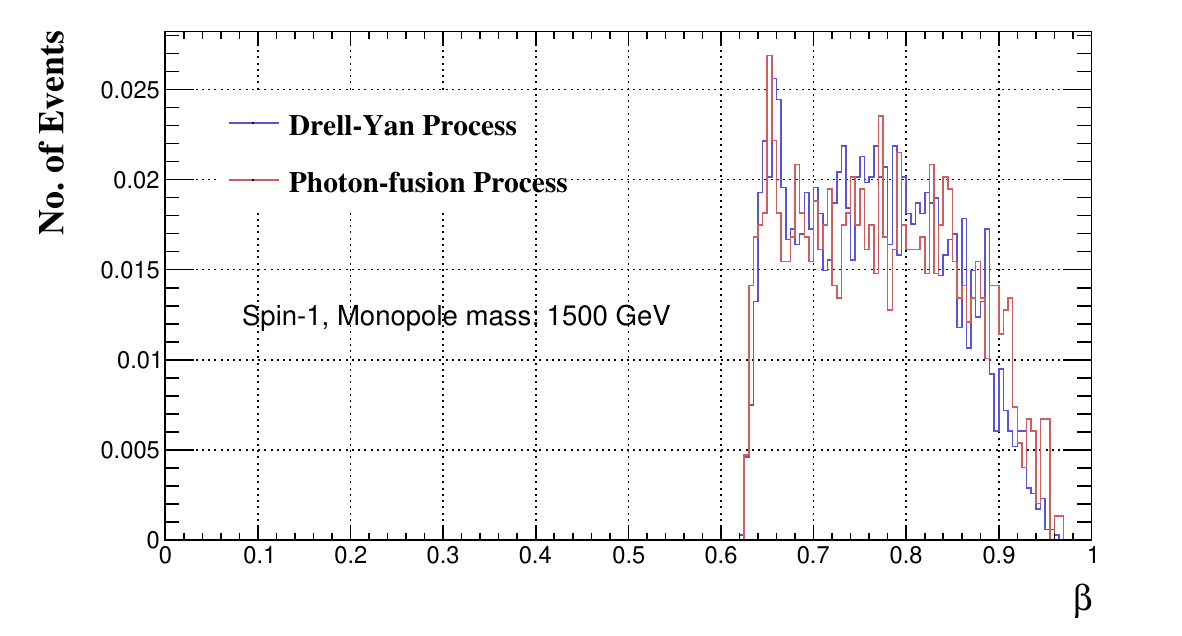}\label{r1}}
    \caption{Comparison of `$\beta$' distribution for DY and PF mechanisms at $\sqrt{s} = 14$ TeV. }
    \label{e4}
\end{figure}

Initial simulations were carried out to analyze the velocity distribution of monopoles ($\beta$), which is vital for their detection in experiments like MoEDAL \cite{MoEDAL014}. The findings are displayed in \autoref{e4}, showcasing a comparative analysis of the two mechanisms under consideration. Note that the velocity $\beta$, calculated in the center-of-mass frame of the colliding protons, is largely dependent on the photon’s probability density function (PDF) in the proton, especially for the PF and DY processes. For scalar monopoles, Fig.~\autoref{l1} indicates that the PF process is expected to produce slower monopoles compared to the DY process, which is beneficial for the detection capabilities of MoEDAL’s Nuclear Track Detectors (NTDs) since they are sensitive to monopoles with low $\beta$. In contrast, for fermionic monopoles, the PF process tends to result in faster monopoles than the DY process, as can be seen in Fig.~\autoref{c1}. Lastly, Fig.~\autoref{r1} suggests that the
$\beta$ distributions for the PF and DY processes are quite comparable.

\begin{figure}[h] 
    \centering
    \subfigure[\,Scalar monopoles]{%
        \includegraphics[width=5.2cm, height=5cm]{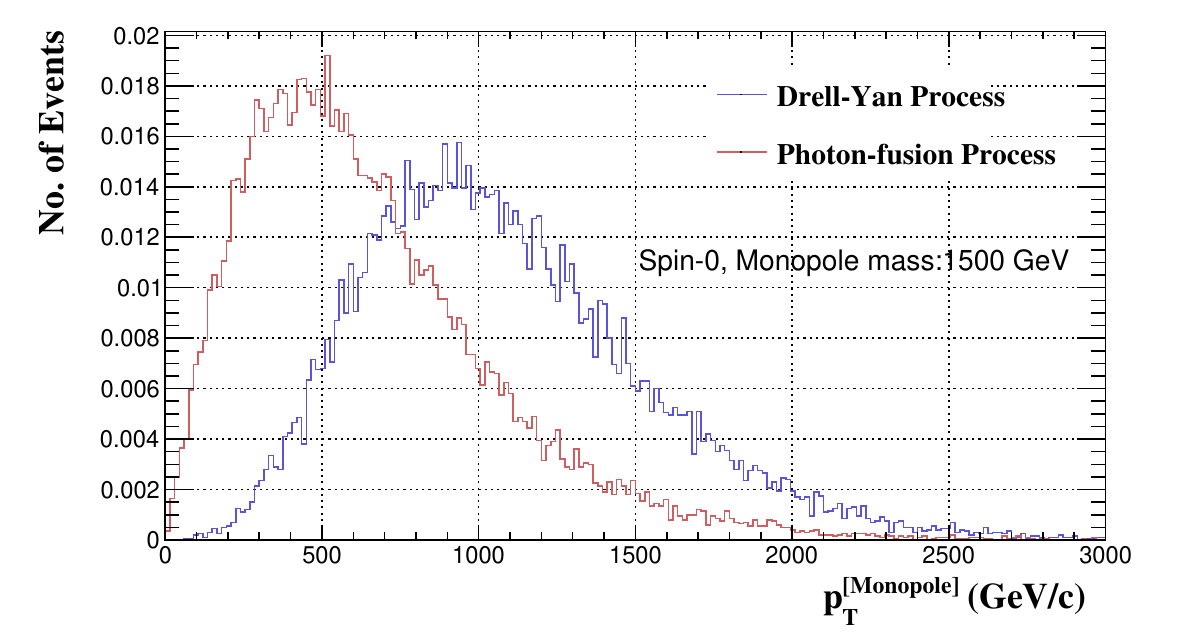}\label{l2}}
    \subfigure[\,Fermionic monopoles]{%
\includegraphics[width=5.2cm,height=5cm]{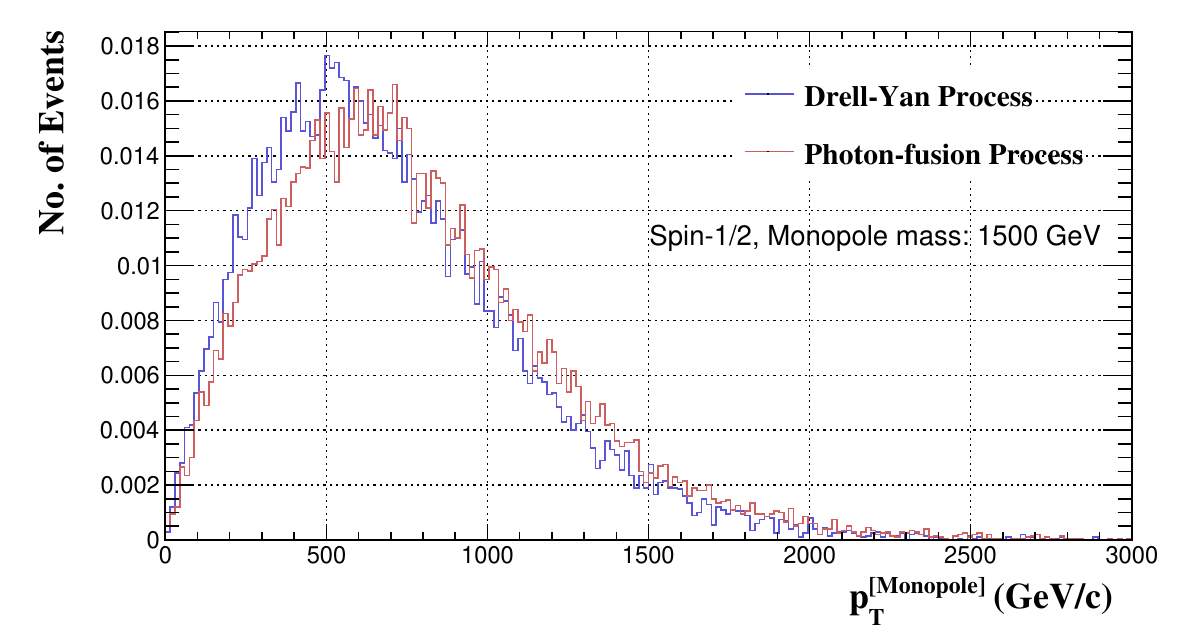}\label{c2}}
    \subfigure[\,Vector monopoles]{%
    \includegraphics[width=5.2cm, height=5cm]{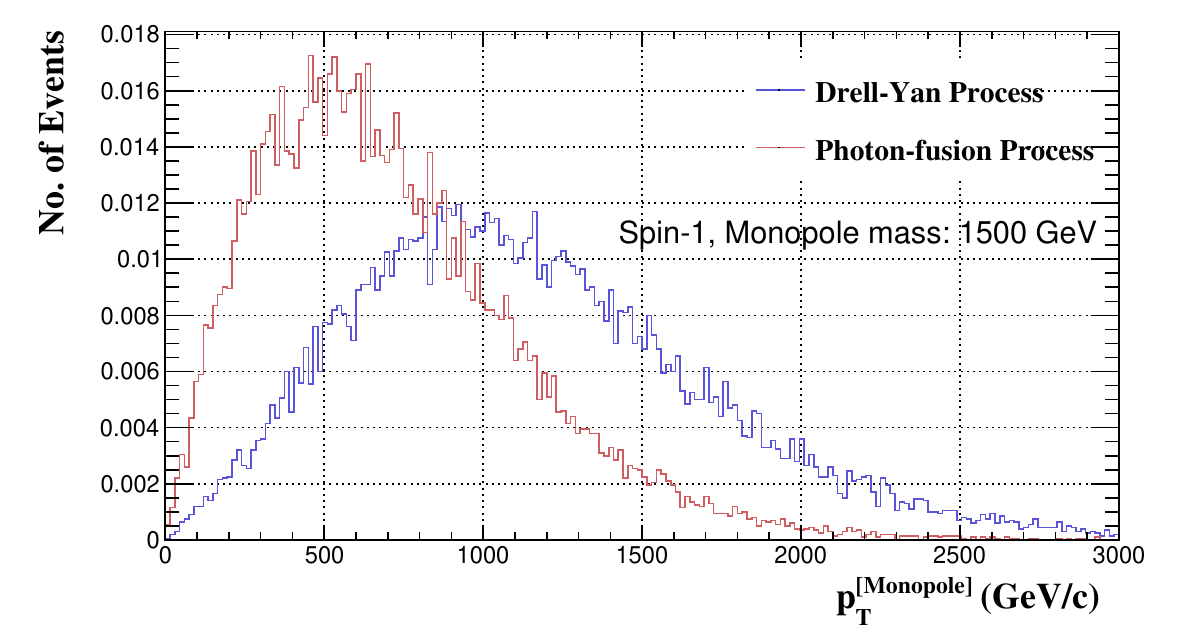}\label{r2}}
    \caption{Comparison of `$p_T$' distribution for DY and PF mechanisms at $\sqrt{s} = 14$ TeV.}
    \label{u1}
\end{figure}

Particularly in pp scattering experiments, the transverse momentum `$p_T$' distribution describes the probability distribution of the momenta of quarks or photons inside a proton, specifically the component perpendicular to the direction of the momentum transfer between the incoming beam and the hadron. Transverse momentum distributions for Drell-Yan and the photon fusion process are shown in \autoref{u1}. For spin $0$, Fig.~\autoref{l2} shows the $p_T$ spectrum of the SM-like condition is noticeably "softer" and its angular dispersion is less central for the photon fusion process than the Drell-Yan. For spin $1/2$ monopole the distributions for both processes are not appreciably different as evident in Fig.~\autoref{c2}. Just like spin 0 cases, in spin $1$ Fig.~\autoref{r2} monopoles the $p_T$ distributions for the Drell-Yan process are more central.  

\begin{figure}[h]     
    \centering
    \subfigure[\,Scalar monopoles]{%
     \includegraphics[width=5.2cm, height=5cm]{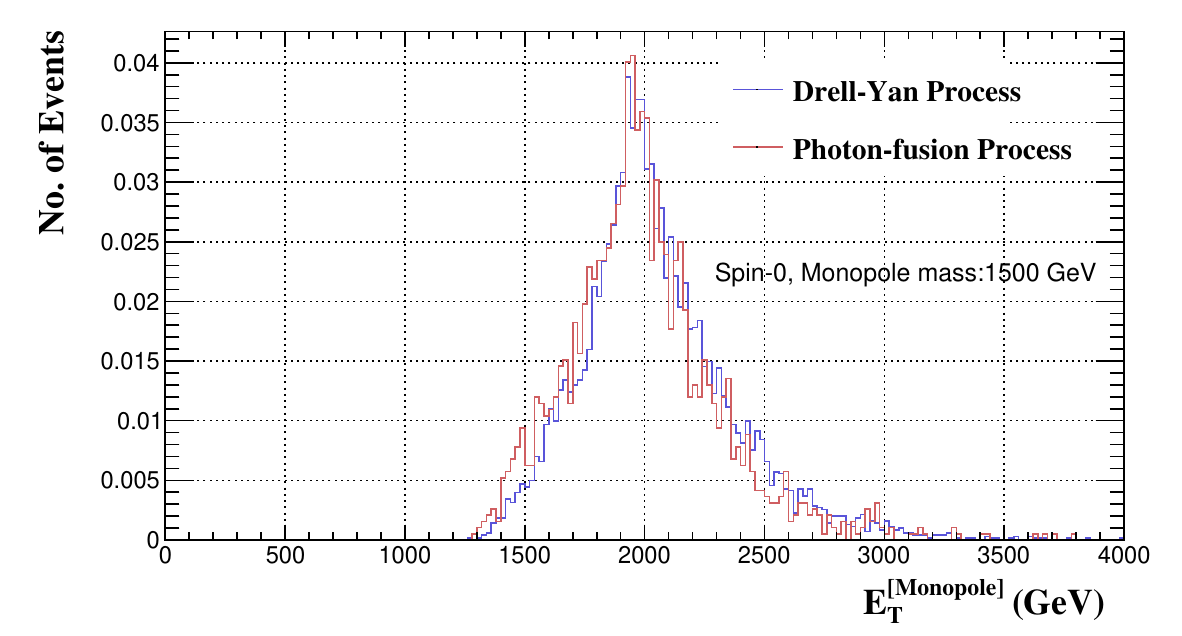}\label{l3}}
\centering
    \subfigure[\,Fermionic monopoles]{%
   \includegraphics[width=5.2cm,height=5cm]{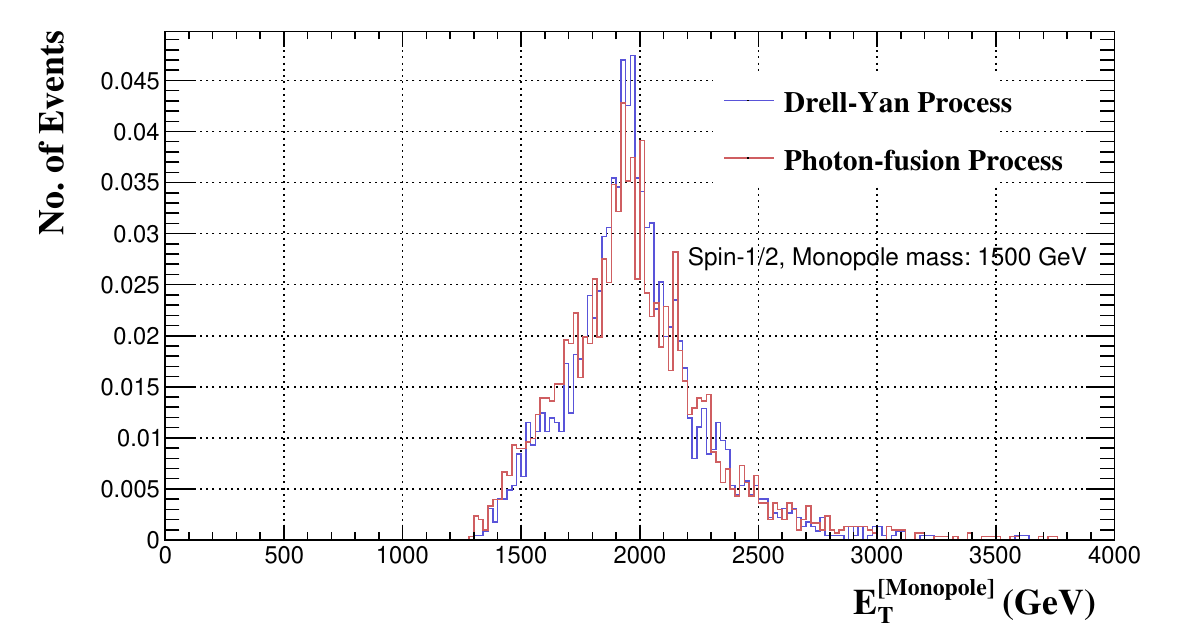}\label{c3}}
   \centering
   \subfigure[\,Vector monopoles]{%
  \includegraphics[width=5.2cm,height=5cm]{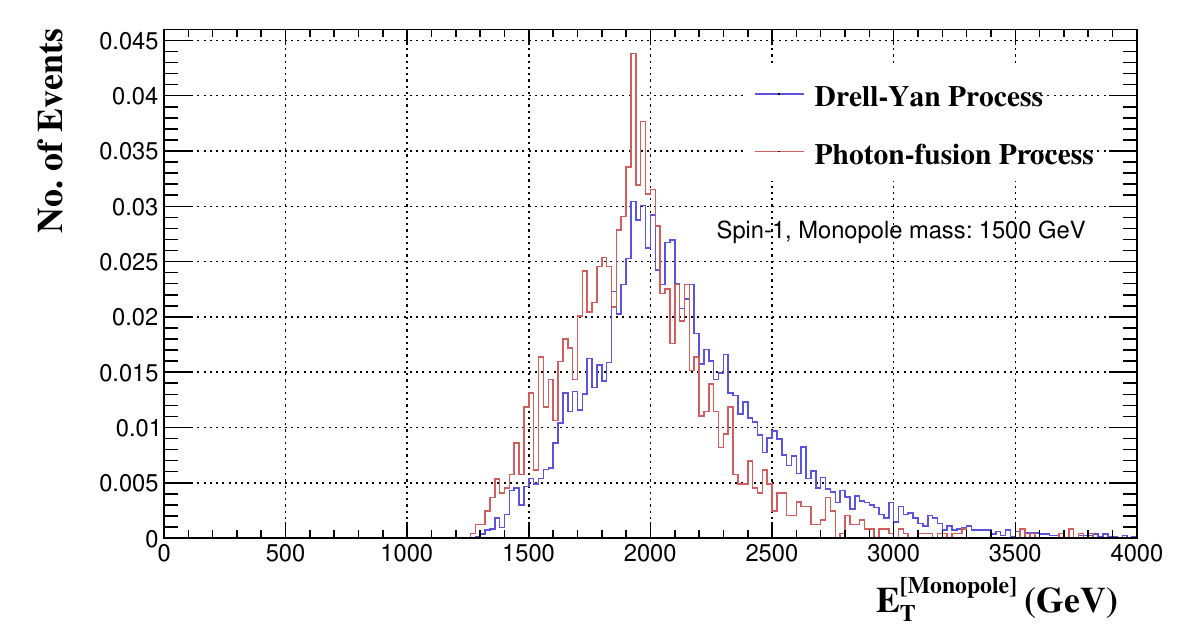}\label{r3}}
    \caption{Comparison of `$E_T$' distributions for DY and PF mechanisms at $\sqrt{s} = 14$ TeV.}
     \label{ef2}
\end{figure}

Transverse energy would be a crucial step in unraveling the properties of magnetic monopoles. Transverse energy distributions are shown in \autoref{ef2}, for both PF and DY for spin 0 (left), 1/2 (central), and 1 (right). The $E_T$ spectrum of the DY process for spin 0 monopoles is almost approaching PF. The $E_T$ distributions of the photon fusion process are more central for spin 1 than the Drell-Yan process.

\begin{figure}[h]  
    \centering
    \subfigure[\,Scalar monopoles with no $k$ term]{%
   \includegraphics[width=5.2cm,height=5cm]{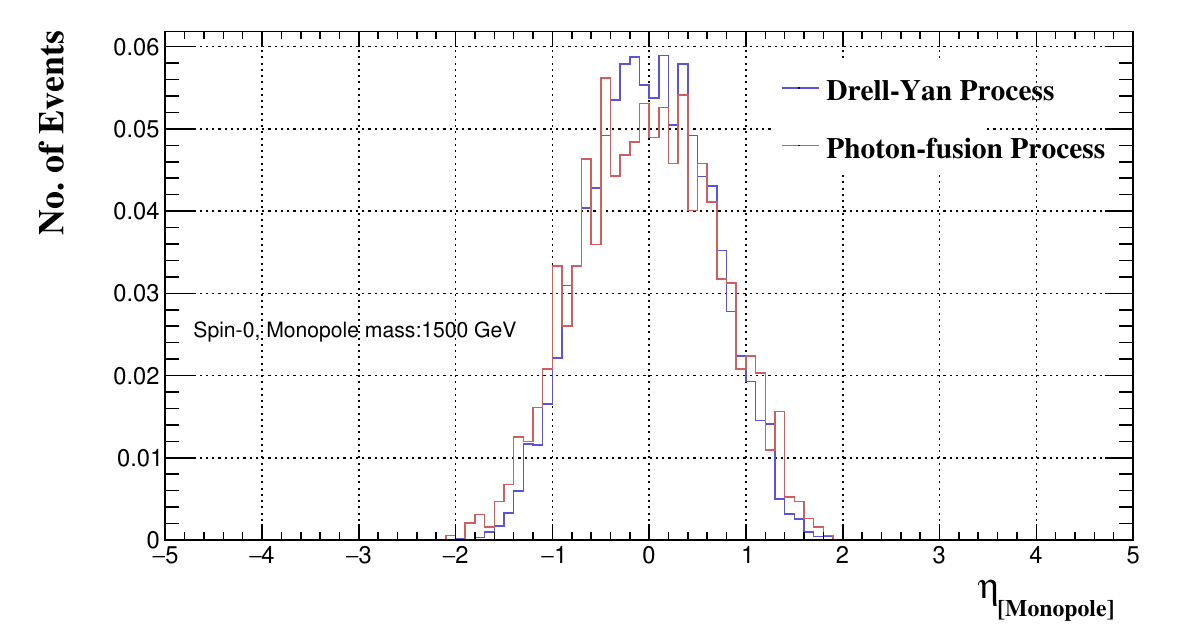}\label{l4}}
 \centering
    \subfigure[\,Fermionic monopoles with $k=0$]{%
   \includegraphics[width=5.2cm,height=5cm]{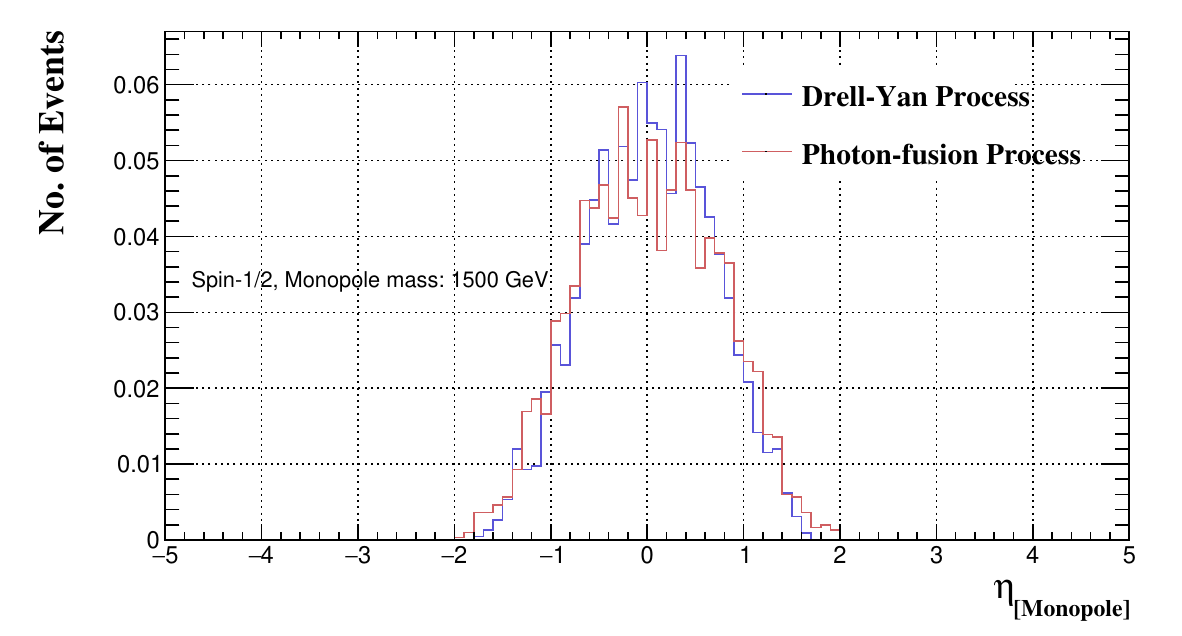}\label{c4}}
   \centering
    \subfigure[\,Vector monopoles with $k=1$]{%
   \includegraphics[width=5.2cm,height=5cm]{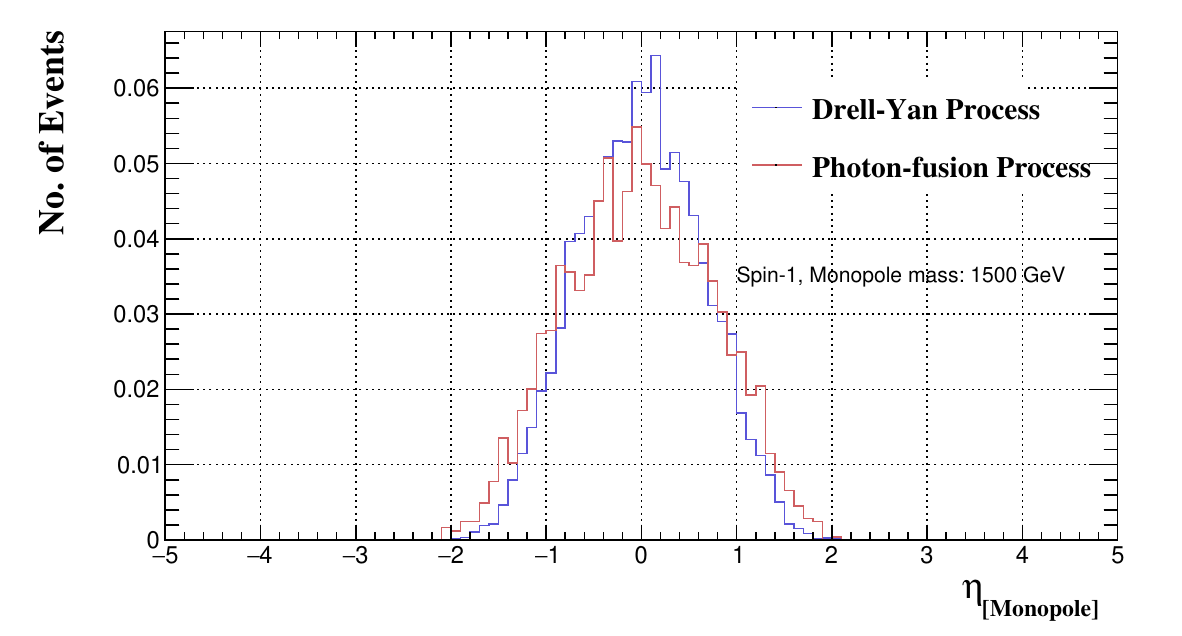}\label{r4}}
   
    \caption{Comparison of `$\eta$' distribution for DY and PF mechanisms at $\sqrt{s} = 14$ TeV.}
    \label{eta1}
\end{figure}

The pseudorapidity `$\eta$' distribution comparison between DY and PF for scalar, fermionic, and vector magnetic monopoles is shown in \autoref{eta1}. Drell-Yan has more central distributions in all three spins.

\begin{figure}[h]  
    \centering
    \begin{subfigure}
     {\includegraphics[width=7cm,height=6cm]{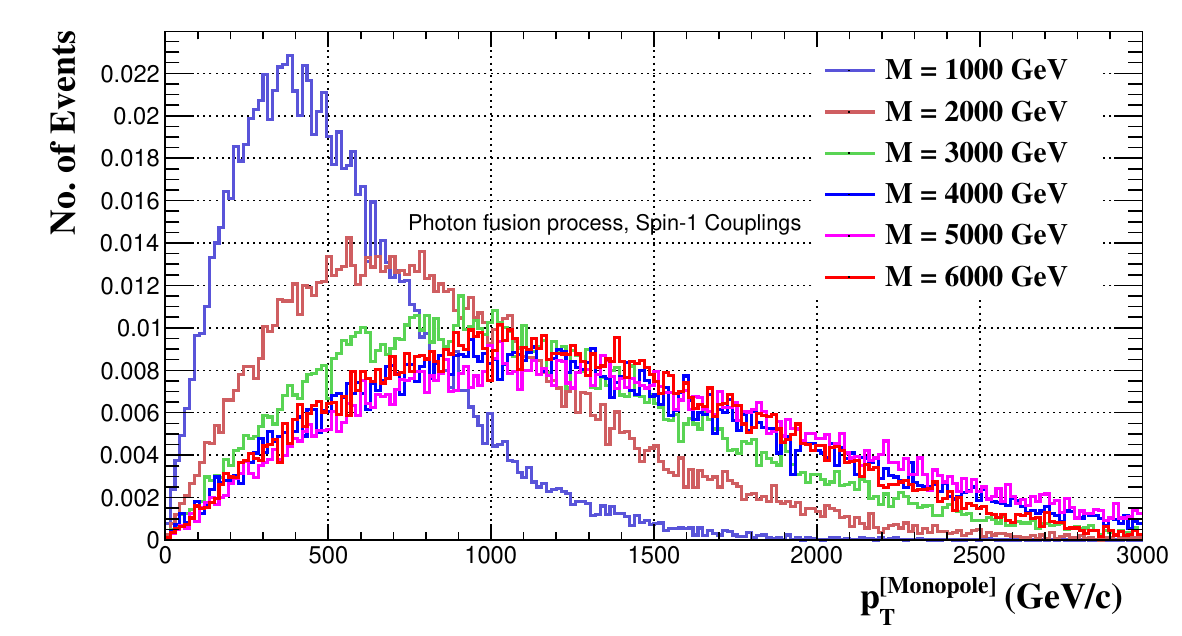}}
 \end{subfigure}
 \begin{subfigure}
     {\includegraphics[width=7cm,height=6cm]{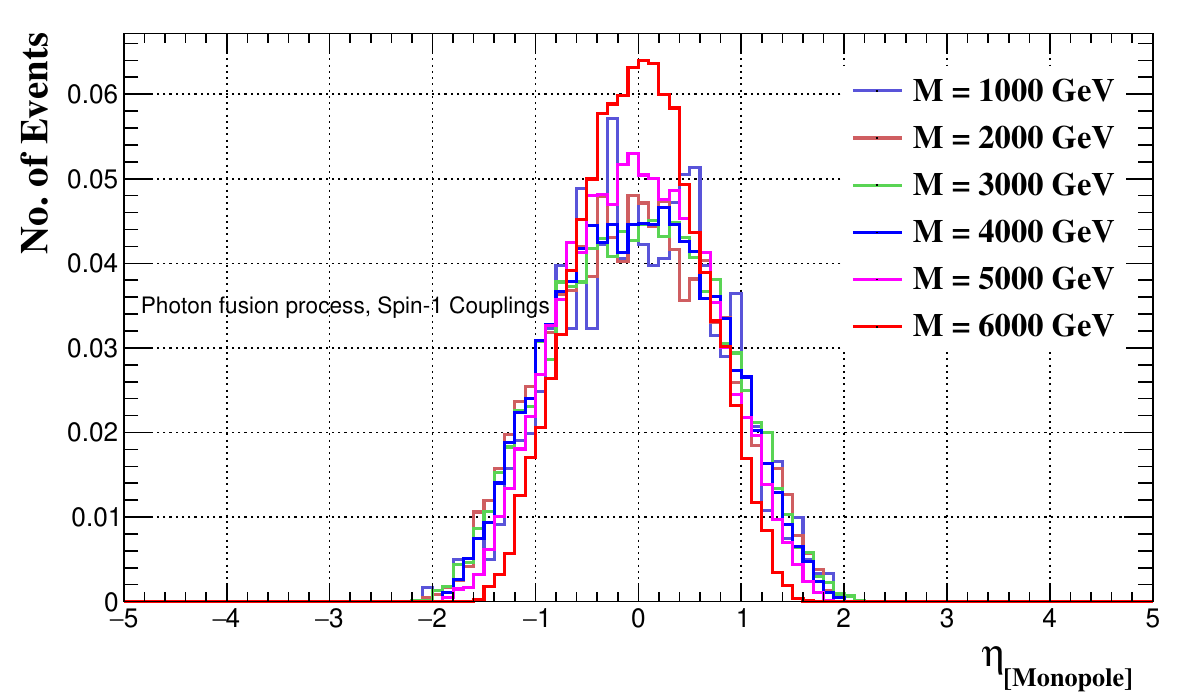}}
 \end{subfigure}
    \caption{The monopole transverse momentum $P_T$ and pseudorapidity $\eta$ distributions for S - 1 for PF process at $\sqrt{s} = 14$ TeV.}
    \label{ptco}
\end{figure}

The comparison of transverse momentum (left) and pseudorapidity distribution (right) at different magnetic monopole masses varying from $1000$ GeV - $6000$ GeV with spin $1$ couplings through photon fusion process is depicted in \autoref{ptco}.

\section{Multivariate Analysis}

The “Toolkit for Multivariate Analysis” (TMVA) is a ROOT-integrated framework \cite{Brun} designed for the parallel processing and evaluation of various multivariate classification methods. It facilitates classification in two event categories: signal and background. While TMVA is primarily developed for high-energy physics (HEP) applications, its utility is not limited to this field alone. It offers suitable preprocessing options for data before it is input into any classifier. TMVA also provides essential data insights, such as correlations between input variables, classifier-specific validations, separation rankings, and powers, as well as efficiency versus background rejection curves for all classifiers.

In our study, we focus on three classifiers: Boosted Decision Trees (BDT), Likelihood, and Multilayer Perceptron (MLP). We base our production samples on Monte Carlo simulations of signal and background processes at an energy of $\sqrt{s},\,=\,100$ TeV. The background process includes jets (QCD), while the signal process involves a monopole pair. Both processes are generated using Pythia-$8$ integrated with \textsc{MadGraph}, and detector simulations are conducted using \textsc{Delphes} \cite{deFavereau}. These signal and background samples are then analyzed using TMVA, which offers a comprehensive suite of multivariate classification algorithms within the ROOT framework.

\subsection{Cut Efficiency of TMVA Methods}

\begin{figure}[h] 
\centering
    \subfigure[\,BDT Output]{%
    \includegraphics[width=8cm,height=6cm]{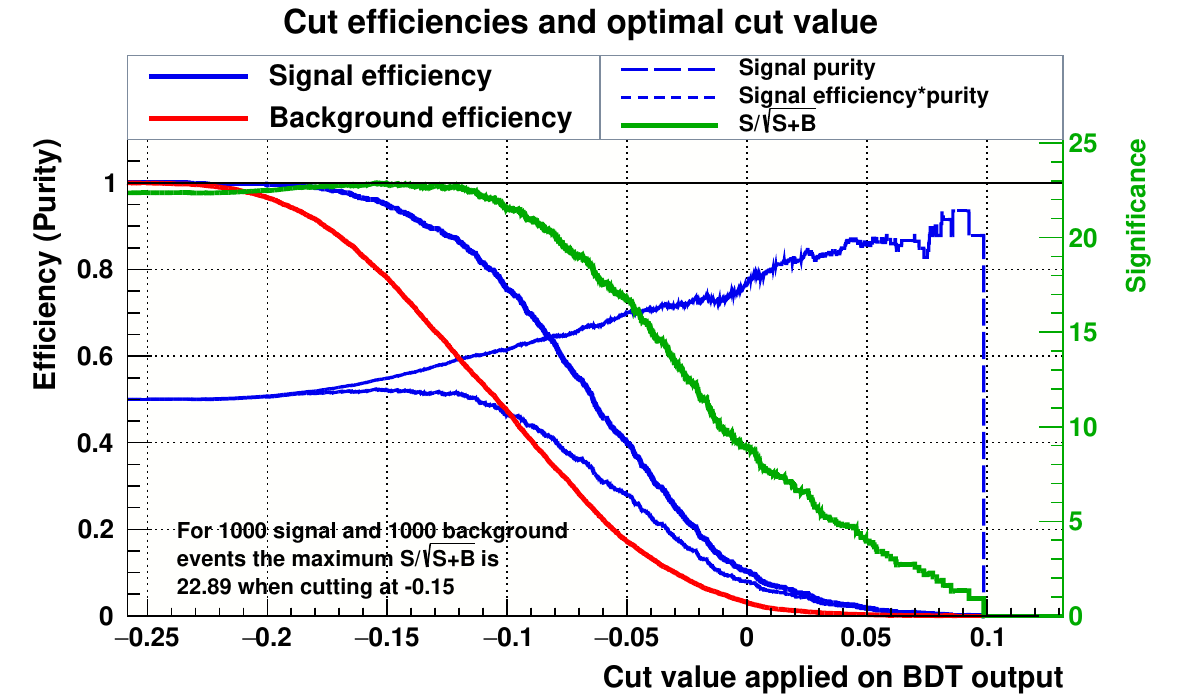}\label{f}}
\centering
    \subfigure[\,MLP Output]{%
    \includegraphics[width=8cm,height=6cm]{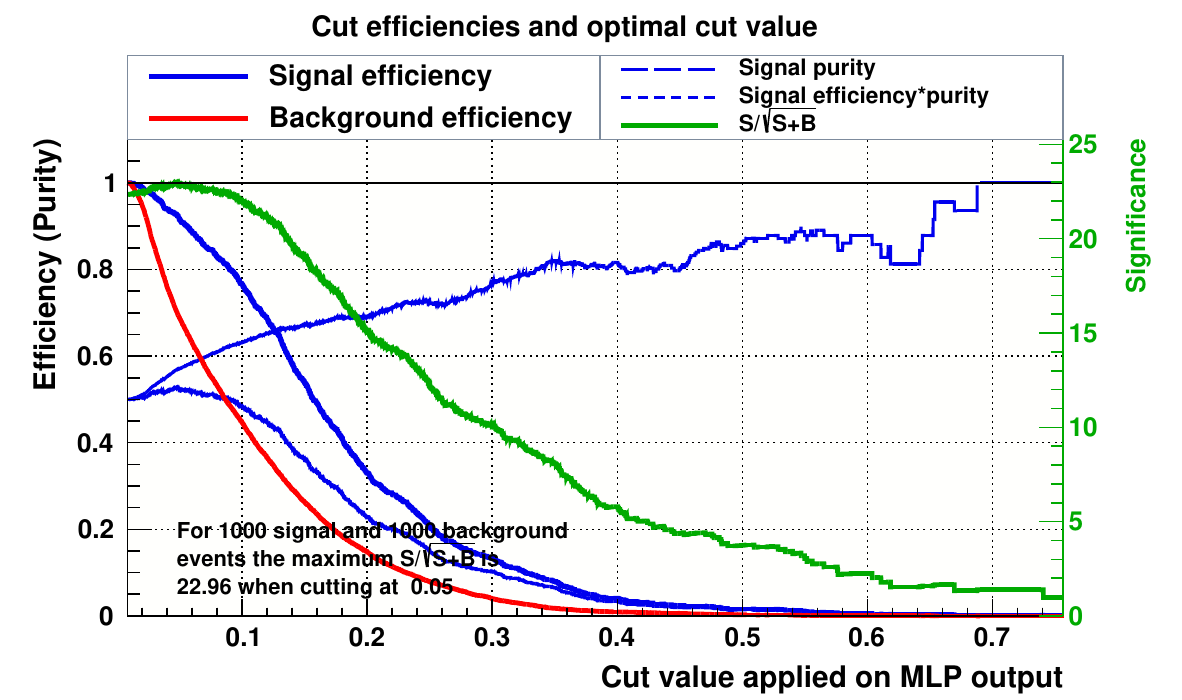}\label{fig:}}
\centering
    \subfigure[\,Likelihood Output]{%
    \includegraphics[width=8cm,height=6cm]{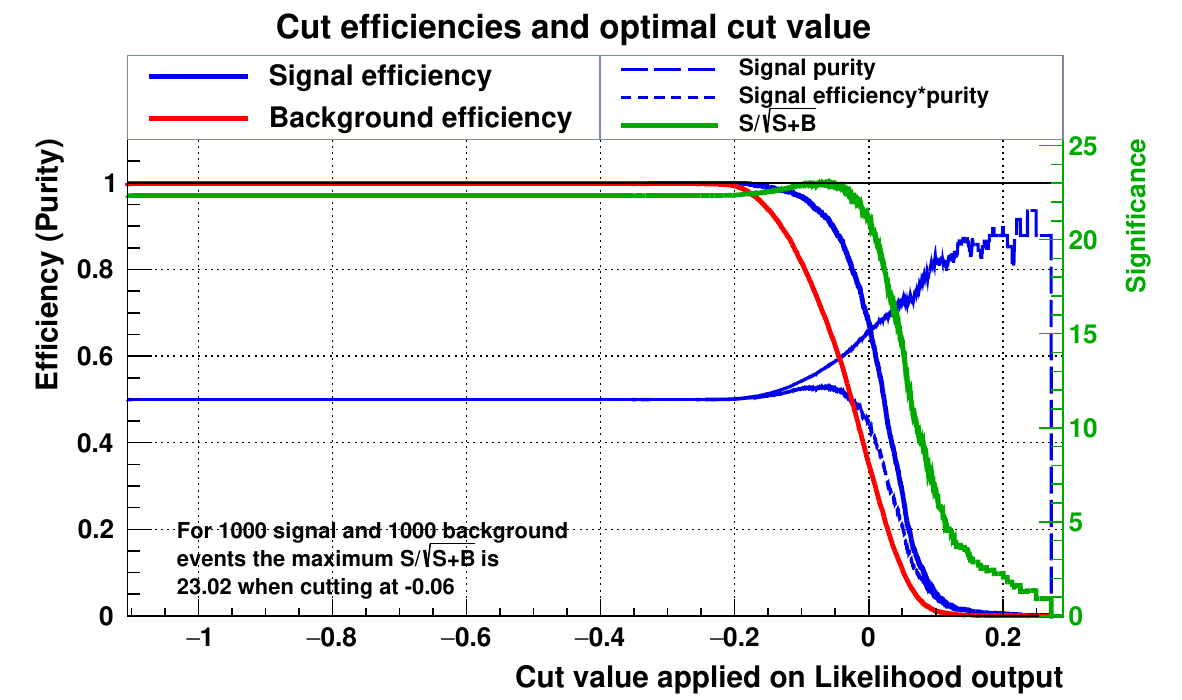}\label{llk}}
    \caption{MVA classifiers signal significance with applied cut values at $\sqrt{s}$ = 100 TeV using Photon fusion process and vector monopoles.}
    \label{tmva}
\end{figure}

 Each classifier has been trained for the $30000$ signal and background events. To check the visibility of the signal process, signal significance is calculated by using cuts as follows:

\begin{equation*}
    \text{Track}.\phi<5  \; \;\:;\:\; \; \text{Track}.\eta<4  \;\; \: ; \;\:\;   E_T^{\text{Missing}}<40 .  
\end{equation*}

Fig.~\autoref{f} shows the signal significance for the BDT classifier by applying a cut value. The signal efficiency is higher in the negative region of the BDT output. Similarly, the $\frac{S}{\sqrt{S+B}}$ is also higher in the negative region with the maximum value of $22.89$ at $-0.15$ with a maximum signal efficiency of $96\%$. The signal efficiency suddenly drops to a value of almost $0.05$.

Similarly, signal significance is calculated for the MLP classifier with applying cuts and is shown in Fig.~\autoref{fig:}. Signal efficiency is a maximum of $2.96$ at the MLP cutting value of $0.05$ with a signal efficiency of $0.9273$. The significance of the
Likelihood in Fig.~\autoref{llk} is $23.02$ at $-0.06$ with signal efficiency $0.9167$.

\begin{table}[h] 
    \centering
    \begin{tabular}{||c|c|c|c|c||}\hline
        \textbf{MVA Classifiers} & \textbf{Optimal-cut} & \textbf{$\frac{S}{\sqrt{S+B}}$}& \textbf{Sig-Eff}&\textbf{Bkg-Eff}  \\\hline
        \textbf{MLP} & 0.05& 22.96&0.9273  & 0.7034\\\hline
        \textbf{Likelihood} &-0.06 &23.02 & 0.9167&0.6693 \\\hline
        \textbf{BDT} &-0.15 &22.89 &0.9617  &0.8034 \\\hline
    \end{tabular}
    \caption{The signal and background efficiency for the number of signal and background.}
    \label{tab:my}
\end{table} 

\subsection{Signal Significance}

\begin{figure}[h] 
    \centering
    {\includegraphics[width=8cm]{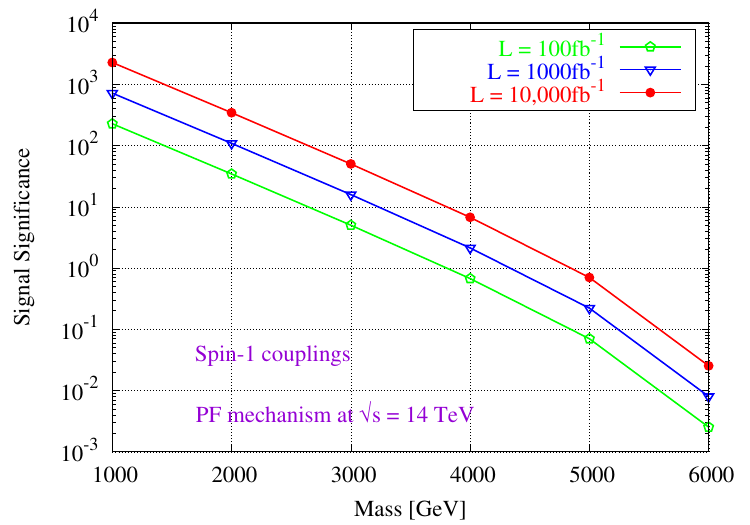}}
    \caption{Comparison of signal significance vs mass across different integrated luminosities at $\sqrt{s}\,=\,14$ TeV and $k\,=\,1$.}
    \label{ss}
\end{figure}

 A higher signal significance (SS) indicates a stronger likelihood that the observed effect is not merely due to random fluctuations or background noise, but rather a genuine discovery. In our study, we focus on the SS in the context of vector monopoles using the PF mechanism at a center-of-mass energy of $14$ TeV. The plot in \autoref{ss} illustrates the relationship between signal significance and the mass of monopoles at different integrated luminosity. It shows that even with the highest amount of data, MMs above $4$ TeV would remain unobservable. 

 The Signal significance or the number of signal-to-background ratios is directly related to the integrated luminosity, which is the total amount of data collected during an experiment. As the integrated luminosity increases, the number of events recorded also increases, which can lead to a higher SS. This is because samples with large statistics allow researchers to identify more and more signal events and better distinguish them from background events.

 \section{Conclusions}

\vphantom{\hphantom{ dealt with the cross-section computation for pair production of monopoles of spin $0,\,1/2,\,1$ through either Photon-Fusion or Drell–Yan processes, also including the magnetic-moment terms. We have employed duality arguments to justify an effective monopole-velocity-dependent magnetic charge in monopole-matter scattering processes. Based on this, we conjecture that such $\beta$-dependent magnetic charges might
also characterize monopole production.\\
 The lack of unitarity and/or renormalisability is restored when the monopole effective theory adopts an SM form, that is when the bare magnetic moment parameter takes on the values $k\,=\,0$ for spin-$1/2$ monopoles, and $k\,=\,1$ for spin-$1$ monopoles.}}
 \vphantom{\hphantom{ The PF cross-section at $\sqrt{s}=14$ TeV is dominant over DY throughout the mass range $1000$ MeV - $6000$ MeV for spin $0,\,1/2,\,\text{and}\,1$. But the cross-section at $\sqrt{s}\,=\,100$ TeV is different, photon fusion is dominant throughout the mass range for spin-1 monopoles, PF remains dominant up to $4000$ MeV but DY dominants $>\,\,4000$ MeV. Similarly, PF is dominant throughout but at $6000$ MeV DY and PF have the same value, which shows that the Drell-Yan dominates for heavy monopoles.\\}}
 
In our examination at the Parton level, prompted by proton-proton collisions at a center-of-mass energy of $14$ TeV, we have investigated the production dynamics of three distinct varieties of magnetic monopoles: scalar, fermionic, and vector types. This analysis encompasses two primary mechanisms, namely the Drell-Yan and the Photon-Fusion processes. We compute the cross-section for magnetic monopole pair production across a spectrum of magnetic moment values, elucidating their respective dependencies. 

We present the kinematic distributions of the transverse momentum, transverse energy, pseudo-rapidity, and monopole boost of magnetic monopoles. Our investigation extends to the reconstruction level, incorporating detector effects and recalculating pseudo-rapidity through a multivariate approach. We employ three machine learning classifiers—BDT, MLP, and Likelihood—to assess the observability of magnetic monopoles and identify optimized cuts that maximize the signal-to-background ratio. Furthermore, we apply various cuts tailored to specific energy and luminosity conditions to enhance observability. Additionally, we compute the signal significance and signal purity to gauge the effectiveness of our analysis. 

The magnetic moment term $k\,= \,0$ for $S-1/2$ and $k\,=\,1$ for $S-1$ added to effective Lagrangian defining the monopole interactions with photons. For these $k$ values absence of unitarity or renormalisability will be restored.

We examine the cross-section of magnetic monopoles across three spin values --- $0, 1/2, \text{and} 1$ --- for both the PF and DY processes at center-of-mass energies of $\sqrt{s} = 12, 13, 14, 27, \text{and} 100$ TeV. The cross-section value for $\sqrt{s}\,=\,14$ TeV is  $1.17\,\times \,10^{-3}$ for spin-$0$ , $2.26\,\times \,10^{-2}$ for spin-$1/2$ and $6.13\,\times\,10^{-2}$ for spin-$1$ monopoles, and for $\sqrt{s}\,=\,100$ TeV  cross section value is  $1.24\,\times\,10^{+4}$ for spin-$0$, $2.36\,\times\,10^{+5}$ for spin-$1/2$ and $6.23\,\times\,10^{+5}$ for spin-$1$ MMs. Notably the cross-section increases at higher CMEs across all spin values. A comparison among spins $0,\,1/2,\,\text{and}\,1$ monopoles in the Drell-Yan process at $\sqrt{s}\,=\,14$ TeV reveals the dominance of vector magnetic monopoles up to $3500$ GeV. Conversely, in the case of Photon-Fusion, vector monopoles exhibit dominance throughout the mass range of $1-6$ TeV, with magnitudes ranging around $\pm 10^8$. This underscores the potential significance of vector monopoles as a strong candidate for further investigation in phenomenological scenarios. 

The cross-section for Photon-Fusion surpasses that of the Drell-Yan processes across all three spin values, maintaining a consistent trend of higher cross-sections at lower masses, which subsequently decrease with increasing mass. This phenomenon can be attributed to the increased energy demand for the production of heavier monopoles at specific magnetic moments.

We conducted a comparative analysis of kinematic distributions essential for experimental investigations, contrasting between the PF and DY production mechanisms for monopoles of spins $0,\,1/2,\,\text{and}\,1$. This approach facilitates the exploration of the photon fusion mechanism, which emerges as dominant at LHC energies. Additionally, based on these kinematic distributions, we delineate the experimental considerations for a perturbatively valid search for monopoles with large magnetic-moment parameters and those exhibiting slow motion. 

Utilizing the Toolkit for Multivariate Analysis (TMVA) classification, we determined that the most effective classifiers for generating magnetic monopoles are MLP, BDT, and likelihood. Through the application of enhanced signal significance and the implementation of tailored cuts to bolster signal efficiency, we achieved optimal performance. Specifically, the BDT classifier exhibited a signal efficiency of $96\%$, while the likelihood classifier demonstrated a signal significance of $23.02$.

\section{Acknowledgement}

The authors thank Arka Santra and Vasiliki A. Mitsou for their invaluable assistance with model files, addressing various model-related issues, and providing guidance on adjusting parameters and effectively utilizing them.

\bibliographystyle{unsrt}
\bibliography{References}

\end{document}